\documentclass[aps,prl,amssymb,twocolumn,showpacs,supersriptaddress,groupeaddress]{revtex4-1}
\usepackage{amssymb}
\usepackage{amsmath}
\usepackage{graphicx}
\usepackage{dcolumn}
\usepackage{bm}
\usepackage{natbib,hyperref}
\usepackage{hyperref}
\usepackage{color}
\hypersetup{colorlinks=true,
    linkcolor=red,
    citecolor=magenta,
    filecolor=green,
    urlcolor=blue
}
\usepackage{float}
\usepackage{sidecap}

\usepackage{multibib}

\graphicspath{{images1/}}

\begin{document}

\title{Quantum emulation of coherent backscattering in a system of superconducting qubits}

\author{Ana Laura Gramajo$^{1,4}$, Dan Campbell$^{1}$, Bharath Kannan$^{1}$, David K. Kim$^{2}$, Alexander Melville$^{2}$, Bethany M. Niedzielski$^{3}$, Jonilyn L. Yoder$^{2}$, Mar\'ia Jos\'e S\'anchez$^{4,5}$, Daniel Dom\'inguez$^{4}$, Simon Gustavsson$^{1}$ and William D. Oliver$^{1,2,3}$\\
}

\affiliation{$^{1}$Research Laboratory of Electronics, Massachusetts Institute of Technology, Cambridge MA 02139, USA}
\affiliation{$^{2}$MIT Lincoln Laboratory, 244 Wood Street, Lexington, MA 02420, USA}
\affiliation{$^{3}$Department of Physics, Massachusetts Institute of Technology, Cambridge MA 02139, USA}
\affiliation{$^{4}$Centro At{\'{o}}mico Bariloche and Instituto Balseiro, 8400 San Carlos de Bariloche, Argentina}
\affiliation{$^{5}$Instituto de Nanociencia y Nanotecnolog\'ia (INN), 8400 San Carlos de Bariloche, Argentina}

\date{\today}
\begin{abstract}
	
In condensed matter systems, coherent backscattering and quantum interference in the presence of time-reversal symmetry lead to well-known phenomena such as weak localization (WL) and universal conductance fluctuations (UCF). Here we use multi-pass Landau-Zener transitions at the avoided crossing of a highly-coherent superconducting qubit to emulate these phenomena.
The average and standard deviation of the qubit transition rate exhibit a dip and peak when the driving waveform is time-reversal symmetric, analogous to WL and UCF, respectively.
The higher coherence of this qubit enabled the realization of both effects, in contrast to earlier work~\cite{gustavsson_2013}, which successfully emulated UCF, but did not observe WL.
This demonstration illustrates the use of non-adiabatic control to implement quantum emulation with superconducting qubits.
\end{abstract}

\maketitle

\sidecaptionvpos{figure}{t}

Studies of mesoscopic disordered structures at cryogenic temperatures exhibit universal phenomena in their electrical conductance arising from the coherent scattering of electrons at random impurities~\cite{abrahams_1979,lee1_1985,altshuler_1988}.
One example is universal conductance fluctuations (UCF)~\cite{webb_1985,lee2_1985,altshuler_1988,benoit_1987}, which are strong fluctuations in the conductance -- on the order of the quantum unit of conductance -- that appear as a function of a parameter, e.g., magnetic field, which effectively alters how the electronic wavefunction samples
the random configuration of scatterers.
Another example is weak localization (WL), a quantum correction to the classical conductance that survives disorder averaging under conditions of
time-reversal symmetry~\cite{datta_1995,ferry_goodnick_1997,bergman_1982}.
The result is a dip in the disordered-averaged conductance (equivalently, a peak in the resistance) at zero magnetic field and when spin-orbit effects are negligible, due to the constructive interference between the symmetric forward- and backward-propagating electron waves arising from impurity scattering~\cite{benoit_1987,washburn_1992}.
In the presence of a magnetic field, the time-reversal symmetry -- and thus the degeneracy in phase evolution -- is lifted for the two paths and the interference leading to the WL effect is abated~\cite{benoit_1987,washburn_1992}.
Studies of WL and UCF provide a method for investigating phenomena related to phase coherence, coherent backscattering, and time-reversal symmetry, and they have been applied to a wide variety of systems ranging from metals~\cite{dolan_1979}, and semiconductors~\cite{bishop_1980} to superconducting solid-state devices~\cite{chen_2014}, quantum dots~\cite{marcus_1992,chan_1985,folk_1996} and graphene~\cite{morozov_2006}, and even for the scattering of light of disordered media~\cite{albada_1985,wolf_1985,scheffold_1998,schreiber_2010}.

This work implements a quantum emulator of WL and UCF phenomena using coherent scattering at an avoided crossing present in coupled superconducting qubits.
The approach is motivated by earlier work in Ref.~\onlinecite{gustavsson_2013}, where an avoided crossing of a single persistent-current flux qubit was used to represent a coherent scattering impurity.
Conceptually, each period of a large-amplitude biharmonic flux signal drives the qubit multiple times through the avoided crossing. Each traversal of the crossing drives the qubit states into quantum superpositions of ground and excited states -- Landau-Zener-Stueckelberg (LZS) transitions -- with output amplitudes related to the size of the avoided crossing, the rate at which the qubit is driven through the crossing, and the resulting quantum interference. The sequence of traversals serve as the scattering sites, and the driven evolution between scattering events accounts for free-evolution phase accumulation of an electron, for example, in a disordered medium. Since the scattering events are imposed by the driving protocol, the time-reversal symmetry (asymmetry) of the system is controlled by the temporal symmetry (asymmetry) of the drive waveform.
\begin{figure*}[ht!]
\centering
	\includegraphics[scale = 0.28]{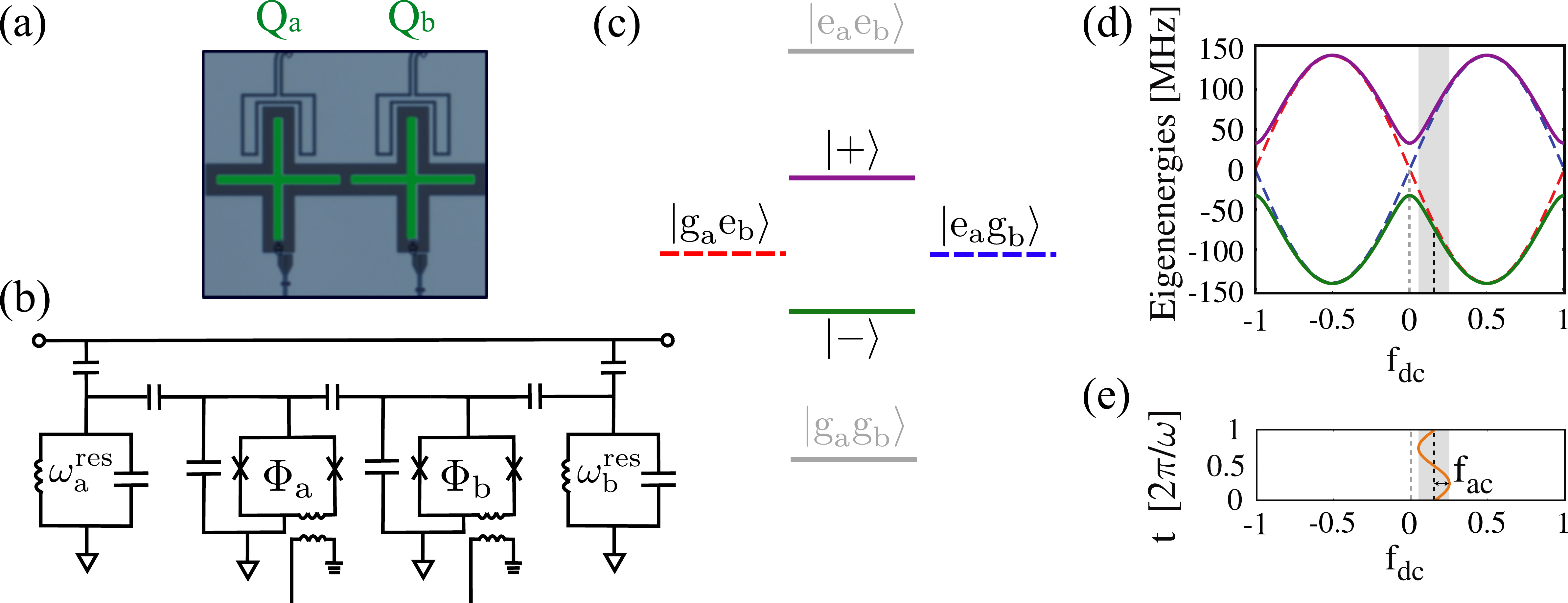}
	\caption{\textbf{Experimental device and energy levels.}
\textbf{(a)} False color micrographe of two capacitively coupled transmon qubits, $\textrm{Q}_{\textrm{a}}$ and $\textrm{Q}_{\textrm{b}}$.
\textbf{(b)} Circuit schematic. $\textrm{Q}_{\textrm{a,b}}$ each have individual control lines used to differentially implement (see text) a static ($f_{\textrm{dc}}$) magnetic flux bias and the biharmonic driving protocol ($f_{\textrm{ac}}$) through the flux biases $\Phi_{\textrm{a,b}}$. Standard microwave gates are used for device initialization, and the qubits are read out by capacitively coupling to individual readout resonators with frequencies $\omega_{\textrm{a,b}}^{\textrm{res}}$.
\textbf{(c)} Energy levels of the two qubits. The red and blue diabatic energies correspond to one excitation of one of the qubits. These levels are frequency tunable using baseband flux control. When degenerate, these levels hybridize to form an avoided crossing of approximately $65.4$ MHz.
\textbf{(d)} Diabatic energies (dashed line) and eigenenergies (solid lines) of the one-excitation manifold $\{ |e_{\text{a}}g_{\text{b}}\rangle,|g_{\text{a}}e_{\text{b}}\rangle \}$ as the function of $f_{dc}$.
\textbf{(e)} Scattering at the avoided crossing is implemented using a driving pulse $f(t) = f_{\textrm{dc}} + f_{\textrm{ac}}(t)$, illustrated here as a sine wave. The driving pulse is applied differentially to each qubit.
}
\label{fig:1}
\end{figure*}

Using this approach, Gustavsson \textit{et al.}~\cite{gustavsson_2013} emulated UCF-type phenomena -- analogous to fluctuations observed in electron transport through a disordered mesoscopic system -- by describing the qubit-state transition rate to electrical conductance. The authors observed fluctuations in the transition rate to the qubit excited state arising from multiple LZS scattering events when measured as a function of the driving waveform asymmetry. However, the analog of WL localization -- a dip in the average transition rate for symmetric driving -- was not observed. Subsequent theoretical work from Ferr\'on \textit{et al.}~\cite{ferron_2017} indicated that both UCF and WL signatures should be observable if the qubit is operated in a higher phase-coherence regime. Indeed, while the niobium qubit used in Ref.~\onlinecite{gustavsson_2013} had a large energy relaxation time ($T_1 \approx 20$ $\mu\textrm{s}$) and a coherence time ($T_2 \approx 20$ $\textrm{ns}$) sufficient to observe LZS interference phenomena including Mach-Zehnder-type interferometry~\cite{Oliver2005,Berns2006}, qubit cooling~\cite{Valenzuela2006}, and amplitude spectroscopy~\cite{Berns2008,Rudner2008}, the phase coherence time was apparently insufficient to observe WL.

In this work, we use coupled aluminum transmon qubits~\cite{koch_2007} to realize a higher-coherence quantum system with a reasonably sized avoided crossing ($\approx 65.4$ MHz). We drive the system using a biharmonic waveform -- with a specified degree of asymmetry -- to emulate electron transport in the presence of multiple scattering events. The resulting transition rate exhibits effects analogous to WL and UCF in its ensemble averaged mean and variance, respectively. The experimental results are in agreement with simulations based on a Floquet formalism.

We utilize an effective two-level system encoded in the single-excitation manifold of two capacitively coupled superconducting transmon qubits~\cite{shim_2016,Campbell_2019} of the X-mon style~\cite{Barends2013} using asymmetric junctions with a 13:1 area ratio~\cite{Hutchings2017}.
The individual transmons $Q_{\textrm{a}}$ and $Q_{\textrm{b}}$ are well-matched
 in frequency,
with maximum frequencies
$\omega_{\textrm{a}}^{\textrm{max}}/2\pi$ = 3.8250 GHz and $\omega_{\textrm{b}}^{\textrm{max}}/2\pi$ = 3.8218 GHz and minimum frequencies
$\omega_{\textrm{a}}^{\textrm{min}}/2\pi$ = 3.5401 GHz and $\omega_{\textrm{b}}^{\textrm{min}}/2\pi$ = 3.5365 GHz, respectively, and they are each frequency-tunable between the two values by a magnetic flux $\Phi_{\textrm{a}}$ and $\Phi_{\textrm{b}}$ (see Fig.~\ref{fig:1}). The qubits are individually coupled to readout resonators at frequencies $\omega_{\textrm{a}}^{\textrm{res}}/2 \pi$ = 7.173262 GHz and $\omega_{\textrm{b}}^{\textrm{res}}/2 \pi$ = 7.203279 GHz, which are used for qubit state discrimination and provide an additional pathway to implement state preparation using microwave gates.

The qubits are each flux-biased at $\Phi_{\textrm{a}} \approx \Phi_{\textrm{b}} \approx 0.25 \Phi_0$, where $\Phi_0=h/2e$ is the superconducting flux quantum, $h$ is Planck's constant, and $e$ is the electron charge.
This bias point is approximately midway between maximum and minimum qubit frequencies such that the uncoupled qubit frequencies $\omega_{\textrm{a}}/2\pi=\omega_{\textrm{b}}/2\pi=3.6809$ GHz are degenerate, leading to the energy level structure
 in the diabatic basis shown in Fig.~\ref{fig:1}c.
Due to the capacitive qubit-qubit coupling, the diabatic states $|g_{\textrm{a}} e_{\textrm{b}} \rangle$ and $|e_{\textrm{a}} g_{\textrm{b}} \rangle$ in the single-excitation manifold hybridize to form the eigenfrequencies shown in Fig.~\ref{fig:1}c. Within this manifold, we can now write an effective two-level system Hamiltonian of the standard form in the basis $\{  |g_{\text{a}} e_{\text{b}}\rangle, |e_{\text{a}} g_{\text{b}} \rangle \}$,
\begin{equation}
	\hat{H}_{\text{eff}}/\hbar =  -\frac{\varepsilon}{2} \hat{{\sigma}}_{z} - \frac{\Delta}{2} \hat{{\sigma}}_{x},
\label{eq:TLS-Hamiltonian}
\end{equation}
where $\hbar$ is the reduced Planck constant $h/2\pi$, $\hat{{\sigma}}_{z}$ and $\hat{{\sigma}}_{x}$ are Pauli matrices, $\varepsilon$ is referenced to the location of the avoided crossing, and $\Delta/ 2\pi = 65.4$ MHz is the transverse coupling strength and, thereby, the size of the avoided crossing.

Excursions about the effective two-level system are driven using a longitudinal flux bias applied differentially to the two qubits,
$\delta \Phi(t) = [\Phi_{\textrm{a}}(t) - \Phi_{\textrm{b}}(t)]/2 \equiv \delta \Phi_{\textrm{dc}} + \delta \Phi_{\textrm{ac}}(t)$,
comprising a time-dependent excursion $\delta \Phi_{\textrm{ac}}$ about a static bias point $\delta \Phi_{\textrm{dc}}$ referenced with respect to the avoided crossing (see Fig.~\ref{fig:1}d and~\ref{fig:1}e).
The drive is parameterized as a unitless reduced flux $f(t) = \delta \Phi(t) / \Phi_0 = f_{\textrm{dc}} + f_{\textrm{ac}}(t)$ by normalizing to the superconducting flux quantum $\Phi_0$~\cite{Oliver2005,Orlando1999}. Due to the large area ratio of the junctions, the diabatic frequency $\varepsilon$ in Eq.~\ref{eq:TLS-Hamiltonian} is approximately sinusoidal~\cite{supp-mat} and represents the response of the system to the drive $f(t)$,
\begin{equation}
	\varepsilon(t) \approx  \delta \omega \sin \left[ 2 \pi f(t) \right],
\label{eq:varepsilon}
\end{equation}
where $\delta \omega = (\bar{\omega}^{\textrm{max}} - \bar{\omega}^{\textrm{min}})/2$, with $\bar{\omega}^{\textrm{max/min}}$ the average of $\omega_{\textrm{a}}^{\textrm{max/min}}$ and $\omega_{\textrm{b}}^{\textrm{max/min}}$, respectively, yielding $\delta \omega / 2\pi = 0.1426$ GHz. The instantaneous frequency of the driven two-level system is $\Omega(t) = \sqrt{\varepsilon^2(t) + \Delta^2}$.

\begin{figure*}[htb!]
\centering
	\includegraphics[scale = 0.28]{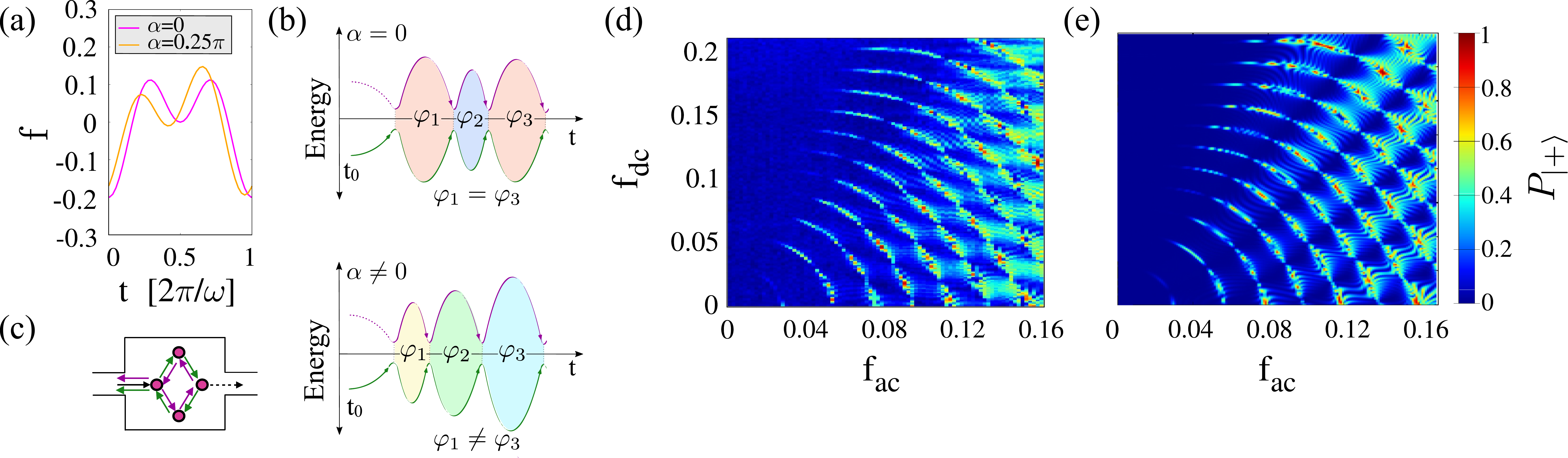}
	\caption{\textbf{Biharmonic driving.}
    \textbf{(a)} Partial period of biharmonic signal that drives the qubit system multiple times through the avoided crossing, for $f_{\textrm{dc}}=0$, $f_{\textrm{ac}}=0.1$, $\alpha = 0$ (magenta line), and $\alpha=0.25\pi$ (orange line), see Eq.(\ref{eq:f(t)}).
    \textbf{(b)} Evolution of system eigenenergy and resulting phase accrual for symmetric ($\alpha=0$) and non-symmetric ($\alpha \neq 0$) biharmonic signals driving the system through the avoided crossing four times. For $\alpha=0$, the system acquires phases in a time-symmetric manner, $\varphi_{1}=\varphi_{3}$. For $\alpha \neq 0$, the phase accrual is no longer time-symmetric.
    \textbf{(c)} Illustration of coherent forward and back-scattering in a disordered condensed matter system.
    \textbf{(d)} Experimental measurement of excited-state ($|+\rangle$) occupation probability (color scale) as the function of $f_{\textrm{dc}}$ and $f_{\textrm{ac}}$, over 20 driving periods, with $\alpha=0$.
    \textbf{(e)} The corresponding numerical simulation.}
	\label{fig:2}
\end{figure*}

To simulate mesoscopic conductance effects, following Ref.~\onlinecite{gustavsson_2013}, we drive the system with a biharmonic signal,
\begin{equation}
\centering
\begin{aligned}
    f(t) = f_{\textrm{dc}} + f_{\textrm{ac}}\left[\cos(\omega t) + \cos(2\omega t+\alpha)\right],
\label{eq:f(t)}
\end{aligned}
\end{equation}
with $\omega / 2 \pi = 10$ MHz, an excursion amplitude $f_{ac}=0.1$, and a phase $\alpha$ that parameterizes the waveform symmetry and thereby the time-reversal symmetry of the system (see Fig.~\ref{fig:2}a).
 As with Ref.~\onlinecite{gustavsson_2013}, the analogy is based on Landau-Zener transitions and the qubit evolution as a phase-space analog of an optical Mach-Zehnder interferometer~\cite{Oliver2005}, where each Landau-Zener transition depicts a scattering event. Fig.~\ref{fig:2}b displays the energy evolution of the qubit during the drive, where each of the interference phases $\varphi = \int \Omega(t) dt$ are given by the shaded area between scattering events.
Although the biharmonic nature of $f(t)$ will drive the system through the avoided crossing up to four times per period for specific values of $f_{\textrm{dc}}$, the accumulated interferences phases $\varphi$ are time-reversal symmetric for $\alpha=0$ (Fig.~\ref{fig:2}b, $\alpha=0$).
 In this case, the qubit trajectories will pick up the same phase during the driven evolution, and they will, therefore, interfere constructively over multiple periods.
However, for $\alpha \neq 0$, the waveform is no longer time-symmetric, and so the interference phases are similarly no longer time-symmetric (Fig.~\ref{fig:2}b, $\alpha=0.25\pi$). The sequential temporal scattering events mimic the spatial scattering in a disordered condensed matter system (Fig.~\ref{fig:2}c).

The control and measurement protocol
consists of the following steps: 1) the qubits are prepared in the two-level system ground-state at flux $f_{\textrm{dc}}$; 2) the drive signal [Eq.(\ref{eq:f(t)})] is applied to the two-level system for an interval of time (a number of periods of the driving field); and 3) the system state is determined by reading out each qubit. Qubit-state readout in the system eigenbases is implemented by adiabatically shifting the qubits away from the avoided crossing region into a region where the uncoupled, diabatic basis states $|g_{\textrm{a}} e_{\textrm{b}} \rangle$ and $|e_{\textrm{a}} g_{\textrm{b}} \rangle$ are essentially identical to the eigenstates. Measuring the individual qubits in this regime are used to infer the occupation probability of the eigenstates~\cite{Chiorescu2003,Bylander2011}.

Using this driving protocol, an analog to the conductance of a mesoscopic system can be emulated by measuring the qubit transition rate $W$, the rate at which population transitions between ground and excited state of the avoided crossing~\cite{gustavsson_2013}.
Multiple sequential passes through a single avoided crossing mimic the scattering amongst a spatial distribution of scatterers (Fig.~\ref{fig:2}c). The phase accumulated between scattering events is dictated by the symmetry and amplitude of the driving waveform.
Unique values of $f_{\textrm{dc}}$ in Eq.(\ref{eq:f(t)}) mimic different scattering configurations, $f_{\textrm{ac}}$ effectively sets transition probabilities and the scattering phases, and the parameter $\alpha$ sets the time-reversal symmetry.
The average transition rate $\langle W\rangle$ -- the analog of average conductance -- is then obtained by ensemble averaging the measured transition rate over all $f_{\textrm{dc}}$ realizations.
See the supplementary information~\cite{supp-mat} for more details.

\begin{figure*}[hbt!]
\centering
	\includegraphics[scale = 0.28]{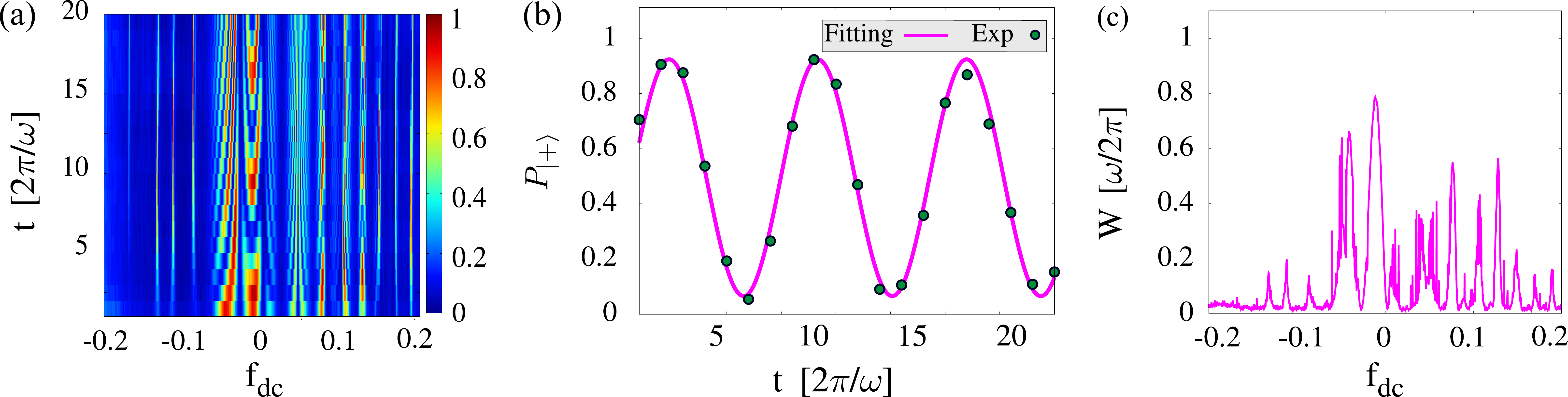}
	\caption{\textbf{Transition rate.}
    \textbf{(a)} Measurement of occupation probability $P_{|+\rangle}$ as a function of time $t$ and bias point $f_{\textrm{dc}}$ with $\alpha=0$.
    \textbf{(b)}  Temporal coherent oscillations in $P_{|+\rangle}$ at $f_{dc}=-0.0126$ due to Landau-Zener-St\"uckelberg transitions at the avoided crossing is fitted (Eq.~\ref{eq:P-sine}) to extract the transition rate $W$ (Eq.~\ref{eq:W}).
    \textbf{(c)} Transition rate $W$ plotted as function of $f_{dc}$ from data in \textbf{(a)}. }
	\label{fig:3}
\end{figure*}

We plot the excited-state population of the two-level system as a function of $f_{\textrm{dc}}$ and  $f_{\textrm{ac}}$ (Fig.~\ref{fig:2}d). The driving field extends for 20 periods with each period being time-symmetric ($\alpha=0$). Each period is approximately $100$ ns, and thus 20 periods is approximatley 2 $\mu$s, smaller than the independently measured coherence times of the two-level system which vary between 4 and 20 $\mu$s, depending on the bias point. Thus, the system remains coherent during the entire driving protocol. A numerical simulation of the time dependent  Schr\"odinger equation (Fig.~\ref{fig:2}e) for these driving parameters is in good agreement with the experimental results~\cite{supp-mat}. The periodic structure observed in Fig.~\ref{fig:2}d and Fig.~\ref{fig:2}e arises from Landau-Zener-St\"uckelberg (LZS) interference upon scattering at the avoided crossings, analogous to a multipass optical interferometer~\cite{Oliver2005,Berns2006,Berns2008,shevchenko_2010}.

In order to obtain the transition rate $W$, we measure the excited-state population $P_{|+\rangle} (t)$ as function of time $t$ and static magnetic flux $f_{\textrm{dc}}$ for a fixed excursion amplitude $f_{\textrm{ac}} = 0.1\Phi_{0}$ and a specific value of asymmetry parameter $\alpha$. As an example, Fig.~\ref{fig:3}a shows the measurements of $P_{|+\rangle} (t)$ for $\alpha=0$. The asymmetry of the resulting excited state population for plus/minus values of $f_{\textrm{dc}}$ results from the driving protocol: at $t=0$, the temporal periodic waveform moves away from $f_{\textrm{dc}}$ in the same flux direction. The first half-period will therefore either approach or move away from the avoided crossing, depending on whether $f_{\textrm{dc}}$ is positive or negative~\cite{Berns2008}. This leads to the left-right asymmetry as a function of $f_{\textrm{dc}}$ in Fig.~\ref{fig:3}a.

\begin{figure*}[hbt!]
    \centering
	\includegraphics[scale=0.25]{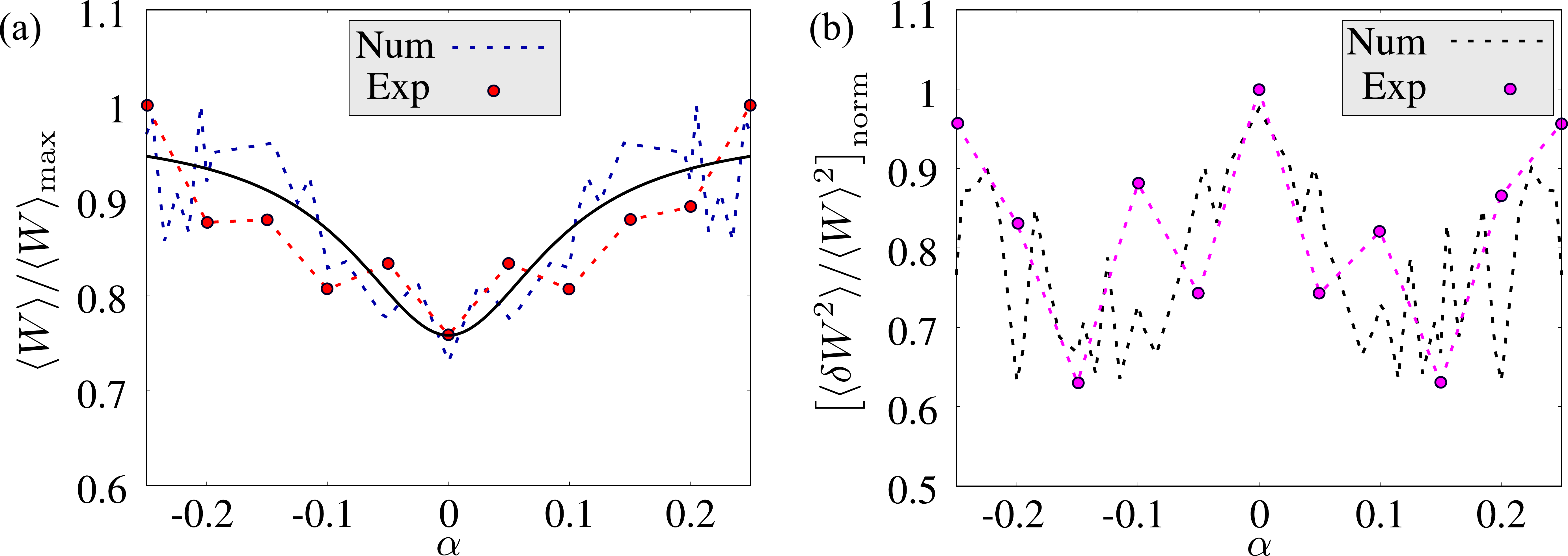}
	\caption{\textbf{First and second-order statistics of the transition rate: WL and UCF.}
    \textbf{(a)} Experimental and numerical results of the normalized transition rate $\langle W \rangle/\langle W \rangle_{\textrm{max}}$ as a function of the asymmetry parameter $\alpha$.
    $\langle W \rangle$ is the transition rate ensemble-averaged over all $f_{dc}$ values (Fig.~\ref{fig:3}c) for fixed $\alpha$, and $\langle W \rangle_{\textrm{max}}$ the corresponding maximum value for each case.
     The bold line is a fit to the data based on the theoretically expected dependence in the WL regime (see text).
    \textbf{(b)} Experimental and numerical results for the normalized variance $\left[ \langle \delta W^2 \rangle / \langle W \rangle^2\right]_{\textrm{norm}} = \left[\left( \langle W^2 \rangle - \langle W \rangle^{2} \right) / \langle W \rangle ^{2}\right]_{\textrm{norm}}$. The normalization $\langle \delta W^2 \rangle / \langle W \rangle^2$ is performed in the same manner as Ref.~\onlinecite{ferron_2017}. There is an extra normalization $[\langle \delta W^2 \rangle / \langle W \rangle^2]_{\textrm{norm}} = [\langle \delta W^2 \rangle / \langle W \rangle^2]/ [\langle \delta W^2 \rangle / \langle W \rangle^2]_{\textrm{max}}$, with $[\langle \delta W^2 \rangle / \langle W \rangle^2]_{\textrm{max}}$ the corresponding maximum value for each case. }
	\label{fig:4}
\end{figure*}

We then fit $P_{|+\rangle} (t)$ for each value of $f_{\textrm{dc}}$ using the function
\begin{equation}
    \begin{aligned}
    	 P_{|+\rangle}(t) = P_{T} \sin( \Omega_{T} t + \varphi) + P_{0},
    \label{eq:P-sine}
    \end{aligned}
\end{equation}
with $P_{T}$, $\Omega_{T}$, $\varphi$ and $P_{0}$ the fitting parameters.  The transition rate can then be computed from the expression:
\begin{equation}
    \begin{aligned}
    	 W(f_{\textrm{dc}}) = \Big|\frac{dP_{|+\rangle}(t)}{dt} \Bigr\rvert_{t=0} \propto \left|P_{T}  \Omega_{T} \right|.
    \label{eq:W}
    \end{aligned}
\end{equation}
The quantity $\left|P_{T}  \Omega_{T} \right|$ serves as a proxy for the transition rate. Fig.~\ref{fig:3}b shows an example fitting of $P_{|+\rangle}(t)$ for $f_{dc} =-0.0126\Phi_{0}$ and $\alpha=0$, from which $W(f_{dc})$ is extracted.  The resulting transition rate $W$ for each value of $f_{dc}$ for $\alpha=0$ is plotted in Fig.~\ref{fig:3}c. Averaging over all values of $f_{\mathrm{dc}}$ (all scattering configurations) leads to the ensemble averaged transition rate $\langle W \rangle$,
\begin{equation}
\begin{aligned}
     \langle W \rangle = \frac{1}{N} \sum_{n=1}^{N} W(f_{\mathrm{dc}}[n])
\label{eq:5}
\end{aligned}
\end{equation}
with $\langle ... \rangle$ the ensemble average over $f_{dc}$, $n$ indexes the values of $f_{dc}$, and $N$ is the total number of $f_{dc}$ values. This procedure is then repeated for different values of symmetry parameter $\alpha$.
The extracted experimental $\langle W \rangle$ for multiple values of $\alpha$ is plotted in Fig.~\ref{fig:4}a, along with results from numerical simulation (see~\cite{supp-mat} for details). Importantly, $\langle W \rangle$ exhibits a dip -- weak localization -- when time-reversal symmetry ($\alpha=0$) is imposed.
 The suppression of the WL correction in the average conductance follows a Lorentzian line-shape with the magnetic field B (in a diffusive transport regime~\cite{Baranger_1993}). Considering that the  parameter $\alpha$ mimics the role of $B$, we plot in Fig.~\ref{fig:4}a  a fit to $\langle  W \rangle_{\alpha} = \tilde{a} - {\tilde{b}}/[{1 + ({\alpha}/{\alpha_{c}})^2}]$, obtaining a good agreement with the experimental results, with $\alpha_c=(0.09\pm0.03)$ (see \cite{supp-mat} for a detailed discussion).

We now proceed to extract the variance in transition rate,
\begin{equation}
    \begin{aligned}
    	\langle \delta W^{2} \rangle = \langle W^2 \rangle - \langle W \rangle^{2}.
        \label{eq:9}
    \end{aligned}
\end{equation}
The experimental and numerical results (Fig.~\ref{fig:4}b) both exhibit a peak in $\langle \delta W^{2} \rangle$ for $\alpha=0$, corresponding to the analog of universal conductance fluctuations (UCF).
We obtain  ${ \langle \delta W^{2} \rangle}/{\langle W \rangle^2}\approx0.6$, which is in quite good agreement with theoretical predictions for disordered systems with many scatterers (see \cite{supp-mat} for further analysis).

An important outcome of this work is the emulation of both WL-type and UCF-type phenomena via coherent scattering at the avoided crossing of a strongly driven qubit system.
Although such UCF-type phenomena were previously reported in Ref.~\onlinecite{gustavsson_2013}, WL was not observed at that time.
The reason was ultimately traced to the relatively short coherence time of the device used in that work, and not an aspect of the driving protocol, as clarified in Ref.~\onlinecite{ferron_2017}.
The present work therefore serves as experimental confirmation of the theory presented in Ref.~\onlinecite{ferron_2017}, and it emphasizes two additional interesting points.
First, even with only a very small number of scattering events, it is possible to emulate behavior that is reminiscent of the well-studied UCF and WL phenomena observed in disordered mesoscopic systems with many more scatterers.
Second, while WL and UCF are both quantum coherent phenomena, WL is apparently more sensitive to quantum coherence in this driven system, requiring a device with higher coherence to manifest itself.

\begin{acknowledgments}
This research was funded by the Office of the Director of National Intelligence (ODNI), Intelligence Advanced Research Projects Activity (IARPA) under Air Force Contract No. FA8721-05-C-0002. The views and conclusions contained herein are those of the authors and should not be interpreted as necessarily representing the official policies or endorsements, either expressed or implied, of ODNI, IARPA, or the US Government. A.L.G, D.D. and M.J.S. are funded by CNEA, CONICET (PIP11220150100756), UNCuyo (P 06/C455) and ANPCyT (PICT2014-1382, PICT2016-0791).
\end{acknowledgments}

\bibliography{references,Review_bibliography_Final}

\begin{thebibliography}{39}%
\makeatletter
\providecommand \@ifxundefined [1]{%
 \@ifx{#1\undefined}
}%
\providecommand \@ifnum [1]{%
 \ifnum #1\expandafter \@firstoftwo
 \else \expandafter \@secondoftwo
 \fi
}%
\providecommand \@ifx [1]{%
 \ifx #1\expandafter \@firstoftwo
 \else \expandafter \@secondoftwo
 \fi
}%
\providecommand \natexlab [1]{#1}%
\providecommand \enquote  [1]{``#1''}%
\providecommand \bibnamefont  [1]{#1}%
\providecommand \bibfnamefont [1]{#1}%
\providecommand \citenamefont [1]{#1}%
\providecommand \href@noop [0]{\@secondoftwo}%
\providecommand \href [0]{\begingroup \@sanitize@url \@href}%
\providecommand \@href[1]{\@@startlink{#1}\@@href}%
\providecommand \@@href[1]{\endgroup#1\@@endlink}%
\providecommand \@sanitize@url [0]{\catcode `\\12\catcode `\$12\catcode
  `\&12\catcode `\#12\catcode `\^12\catcode `\_12\catcode `\%12\relax}%
\providecommand \@@startlink[1]{}%
\providecommand \@@endlink[0]{}%
\providecommand \url  [0]{\begingroup\@sanitize@url \@url }%
\providecommand \@url [1]{\endgroup\@href {#1}{\urlprefix }}%
\providecommand \urlprefix  [0]{URL }%
\providecommand \Eprint [0]{\href }%
\providecommand \doibase [0]{http://dx.doi.org/}%
\providecommand \selectlanguage [0]{\@gobble}%
\providecommand \bibinfo  [0]{\@secondoftwo}%
\providecommand \bibfield  [0]{\@secondoftwo}%
\providecommand \translation [1]{[#1]}%
\providecommand \BibitemOpen [0]{}%
\providecommand \bibitemStop [0]{}%
\providecommand \bibitemNoStop [0]{.\EOS\space}%
\providecommand \EOS [0]{\spacefactor3000\relax}%
\providecommand \BibitemShut  [1]{\csname bibitem#1\endcsname}%
\let\auto@bib@innerbib\@empty
\bibitem [{\citenamefont {Gustavsson}\ \emph {et~al.}(2013)\citenamefont
  {Gustavsson}, \citenamefont {Bylander},\ and\ \citenamefont
  {Oliver}}]{gustavsson_2013}%
  \BibitemOpen
  \bibfield  {author} {\bibinfo {author} {\bibfnamefont {S.}~\bibnamefont
  {Gustavsson}}, \bibinfo {author} {\bibfnamefont {J.}~\bibnamefont
  {Bylander}}, \ and\ \bibinfo {author} {\bibfnamefont {W.~D.}\ \bibnamefont
  {Oliver}},\ }\href {\doibase 10.1103/PhysRevLett.110.016603} {\bibfield
  {journal} {\bibinfo  {journal} {Phys. Rev. Lett.}\ }\textbf {\bibinfo
  {volume} {110}},\ \bibinfo {pages} {016603} (\bibinfo {year}
  {2013})}\BibitemShut {NoStop}%
\bibitem [{\citenamefont {Abrahams}\ \emph {et~al.}(1979)\citenamefont
  {Abrahams}, \citenamefont {Anderson}, \citenamefont {Licciardello},\ and\
  \citenamefont {Ramakrishnan}}]{abrahams_1979}%
  \BibitemOpen
  \bibfield  {author} {\bibinfo {author} {\bibfnamefont {E.}~\bibnamefont
  {Abrahams}}, \bibinfo {author} {\bibfnamefont {P.~W.}\ \bibnamefont
  {Anderson}}, \bibinfo {author} {\bibfnamefont {D.~C.}\ \bibnamefont
  {Licciardello}}, \ and\ \bibinfo {author} {\bibfnamefont {T.~V.}\
  \bibnamefont {Ramakrishnan}},\ }\href {\doibase 10.1103/PhysRevLett.42.673}
  {\bibfield  {journal} {\bibinfo  {journal} {Phys. Rev. Lett.}\ }\textbf
  {\bibinfo {volume} {42}},\ \bibinfo {pages} {673} (\bibinfo {year}
  {1979})}\BibitemShut {NoStop}%
\bibitem [{\citenamefont {Lee}\ and\ \citenamefont
  {Ramakrishnan}(1985)}]{lee1_1985}%
  \BibitemOpen
  \bibfield  {author} {\bibinfo {author} {\bibfnamefont {P.~A.}\ \bibnamefont
  {Lee}}\ and\ \bibinfo {author} {\bibfnamefont {T.~V.}\ \bibnamefont
  {Ramakrishnan}},\ }\href {\doibase 10.1103/RevModPhys.57.287} {\bibfield
  {journal} {\bibinfo  {journal} {Rev. Mod. Phys.}\ }\textbf {\bibinfo {volume}
  {57}},\ \bibinfo {pages} {287} (\bibinfo {year} {1985})}\BibitemShut
  {NoStop}%
\bibitem [{\citenamefont {Al'tshuler}\ and\ \citenamefont
  {Lee}(1988)}]{altshuler_1988}%
  \BibitemOpen
  \bibfield  {author} {\bibinfo {author} {\bibfnamefont {B.~L.}\ \bibnamefont
  {Al'tshuler}}\ and\ \bibinfo {author} {\bibfnamefont {P.~A.}\ \bibnamefont
  {Lee}},\ }\href {\doibase 10.1063/1.881139} {\bibfield  {journal} {\bibinfo
  {journal} {Physics Today}\ }\textbf {\bibinfo {volume} {41}},\ \bibinfo
  {pages} {36} (\bibinfo {year} {1988})}\BibitemShut {NoStop}%
\bibitem [{\citenamefont {Webb}\ \emph {et~al.}(1985)\citenamefont {Webb},
  \citenamefont {Washburn}, \citenamefont {Umbach},\ and\ \citenamefont
  {Laibowitz}}]{webb_1985}%
  \BibitemOpen
  \bibfield  {author} {\bibinfo {author} {\bibfnamefont {R.~A.}\ \bibnamefont
  {Webb}}, \bibinfo {author} {\bibfnamefont {S.}~\bibnamefont {Washburn}},
  \bibinfo {author} {\bibfnamefont {C.~P.}\ \bibnamefont {Umbach}}, \ and\
  \bibinfo {author} {\bibfnamefont {R.~B.}\ \bibnamefont {Laibowitz}},\ }\href
  {\doibase 10.1103/PhysRevLett.54.2696} {\bibfield  {journal} {\bibinfo
  {journal} {Phys. Rev. Lett.}\ }\textbf {\bibinfo {volume} {54}},\ \bibinfo
  {pages} {2696} (\bibinfo {year} {1985})}\BibitemShut {NoStop}%
\bibitem [{\citenamefont {Lee}\ and\ \citenamefont {Stone}(1985)}]{lee2_1985}%
  \BibitemOpen
  \bibfield  {author} {\bibinfo {author} {\bibfnamefont {P.~A.}\ \bibnamefont
  {Lee}}\ and\ \bibinfo {author} {\bibfnamefont {A.~D.}\ \bibnamefont
  {Stone}},\ }\href {\doibase 10.1103/PhysRevLett.55.1622} {\bibfield
  {journal} {\bibinfo  {journal} {Phys. Rev. Lett.}\ }\textbf {\bibinfo
  {volume} {55}},\ \bibinfo {pages} {1622} (\bibinfo {year}
  {1985})}\BibitemShut {NoStop}%
\bibitem [{\citenamefont {Benoit}\ \emph {et~al.}(1987)\citenamefont {Benoit},
  \citenamefont {Umbach}, \citenamefont {Laibowitz},\ and\ \citenamefont
  {Webb}}]{benoit_1987}%
  \BibitemOpen
  \bibfield  {author} {\bibinfo {author} {\bibfnamefont {A.}~\bibnamefont
  {Benoit}}, \bibinfo {author} {\bibfnamefont {C.~P.}\ \bibnamefont {Umbach}},
  \bibinfo {author} {\bibfnamefont {R.~B.}\ \bibnamefont {Laibowitz}}, \ and\
  \bibinfo {author} {\bibfnamefont {R.~A.}\ \bibnamefont {Webb}},\ }\href
  {\doibase 10.1103/PhysRevLett.58.2343} {\bibfield  {journal} {\bibinfo
  {journal} {Phys. Rev. Lett.}\ }\textbf {\bibinfo {volume} {58}},\ \bibinfo
  {pages} {2343} (\bibinfo {year} {1987})}\BibitemShut {NoStop}%
\bibitem [{\citenamefont {Datta}(1995)}]{datta_1995}%
  \BibitemOpen
  \bibfield  {author} {\bibinfo {author} {\bibfnamefont {S.}~\bibnamefont
  {Datta}},\ }\href {\doibase 10.1017/CBO9780511805776} {\emph {\bibinfo
  {title} {Electronic Transport in Mesoscopic Systems}}},\ Cambridge Studies in
  Semiconductor Physics and Microelectronic Engineering\ (\bibinfo  {publisher}
  {Cambridge University Press},\ \bibinfo {year} {1995})\BibitemShut {NoStop}%
\bibitem [{\citenamefont {Ferry}\ and\ \citenamefont
  {Goodnick}(1997)}]{ferry_goodnick_1997}%
  \BibitemOpen
  \bibfield  {author} {\bibinfo {author} {\bibfnamefont {D.}~\bibnamefont
  {Ferry}}\ and\ \bibinfo {author} {\bibfnamefont {S.~M.}\ \bibnamefont
  {Goodnick}},\ }\href {\doibase 10.1017/CBO9780511626128} {\emph {\bibinfo
  {title} {Transport in Nanostructures}}},\ Cambridge Studies in Semiconductor
  Physics and Microelectronic Engineering\ (\bibinfo  {publisher} {Cambridge
  University Press},\ \bibinfo {year} {1997})\BibitemShut {NoStop}%
\bibitem [{\citenamefont {Bergmann}(1982)}]{bergman_1982}%
  \BibitemOpen
  \bibfield  {author} {\bibinfo {author} {\bibfnamefont {G.}~\bibnamefont
  {Bergmann}},\ }\href {\doibase 10.1103/PhysRevB.25.2937} {\bibfield
  {journal} {\bibinfo  {journal} {Phys. Rev. B}\ }\textbf {\bibinfo {volume}
  {25}},\ \bibinfo {pages} {2937} (\bibinfo {year} {1982})}\BibitemShut
  {NoStop}%
\bibitem [{\citenamefont {Washburn}\ and\ \citenamefont
  {Webb}(1992)}]{washburn_1992}%
  \BibitemOpen
  \bibfield  {author} {\bibinfo {author} {\bibfnamefont {S.}~\bibnamefont
  {Washburn}}\ and\ \bibinfo {author} {\bibfnamefont {R.~A.}\ \bibnamefont
  {Webb}},\ }\href {\doibase 10.1088/0034-4885/55/8/004} {\bibfield  {journal}
  {\bibinfo  {journal} {Reports on Progress in Physics}\ }\textbf {\bibinfo
  {volume} {55}},\ \bibinfo {pages} {1311} (\bibinfo {year}
  {1992})}\BibitemShut {NoStop}%
\bibitem [{\citenamefont {Dolan}\ and\ \citenamefont
  {Osheroff}(1979)}]{dolan_1979}%
  \BibitemOpen
  \bibfield  {author} {\bibinfo {author} {\bibfnamefont {G.~J.}\ \bibnamefont
  {Dolan}}\ and\ \bibinfo {author} {\bibfnamefont {D.~D.}\ \bibnamefont
  {Osheroff}},\ }\href {\doibase 10.1103/PhysRevLett.43.721} {\bibfield
  {journal} {\bibinfo  {journal} {Phys. Rev. Lett.}\ }\textbf {\bibinfo
  {volume} {43}},\ \bibinfo {pages} {721} (\bibinfo {year} {1979})}\BibitemShut
  {NoStop}%
\bibitem [{\citenamefont {Bishop}\ \emph {et~al.}(1980)\citenamefont {Bishop},
  \citenamefont {Tsui},\ and\ \citenamefont {Dynes}}]{bishop_1980}%
  \BibitemOpen
  \bibfield  {author} {\bibinfo {author} {\bibfnamefont {D.~J.}\ \bibnamefont
  {Bishop}}, \bibinfo {author} {\bibfnamefont {D.~C.}\ \bibnamefont {Tsui}}, \
  and\ \bibinfo {author} {\bibfnamefont {R.~C.}\ \bibnamefont {Dynes}},\ }\href
  {\doibase 10.1103/PhysRevLett.44.1153} {\bibfield  {journal} {\bibinfo
  {journal} {Phys. Rev. Lett.}\ }\textbf {\bibinfo {volume} {44}},\ \bibinfo
  {pages} {1153} (\bibinfo {year} {1980})}\BibitemShut {NoStop}%
\bibitem [{\citenamefont {Chen}\ \emph {et~al.}(2014)\citenamefont {Chen},
  \citenamefont {Roushan}, \citenamefont {Sank}, \citenamefont {Neill},
  \citenamefont {Lucero}, \citenamefont {Mariantoni}, \citenamefont {Barends},
  \citenamefont {Chiaro}, \citenamefont {Kelly}, \citenamefont {Megrant},
  \citenamefont {Mutus}, \citenamefont {O'Malley}, \citenamefont {Vainsencher},
  \citenamefont {Wenner}, \citenamefont {White}, \citenamefont {Yin},
  \citenamefont {Cleland},\ and\ \citenamefont {Martinis}}]{chen_2014}%
  \BibitemOpen
  \bibfield  {author} {\bibinfo {author} {\bibfnamefont {Y.}~\bibnamefont
  {Chen}}, \bibinfo {author} {\bibfnamefont {P.}~\bibnamefont {Roushan}},
  \bibinfo {author} {\bibfnamefont {D.}~\bibnamefont {Sank}}, \bibinfo {author}
  {\bibfnamefont {C.}~\bibnamefont {Neill}}, \bibinfo {author} {\bibfnamefont
  {E.}~\bibnamefont {Lucero}}, \bibinfo {author} {\bibfnamefont
  {M.}~\bibnamefont {Mariantoni}}, \bibinfo {author} {\bibfnamefont
  {R.}~\bibnamefont {Barends}}, \bibinfo {author} {\bibfnamefont
  {B.}~\bibnamefont {Chiaro}}, \bibinfo {author} {\bibfnamefont
  {J.}~\bibnamefont {Kelly}}, \bibinfo {author} {\bibfnamefont
  {A.}~\bibnamefont {Megrant}}, \bibinfo {author} {\bibfnamefont {J.~Y.}\
  \bibnamefont {Mutus}}, \bibinfo {author} {\bibfnamefont {P.~J.~J.}\
  \bibnamefont {O'Malley}}, \bibinfo {author} {\bibfnamefont {A.}~\bibnamefont
  {Vainsencher}}, \bibinfo {author} {\bibfnamefont {J.}~\bibnamefont {Wenner}},
  \bibinfo {author} {\bibfnamefont {T.~C.}\ \bibnamefont {White}}, \bibinfo
  {author} {\bibfnamefont {Y.}~\bibnamefont {Yin}}, \bibinfo {author}
  {\bibfnamefont {A.~N.}\ \bibnamefont {Cleland}}, \ and\ \bibinfo {author}
  {\bibfnamefont {J.~M.}\ \bibnamefont {Martinis}},\ }\href {\doibase
  10.1038/ncomms6184} {\bibfield  {journal} {\bibinfo  {journal} {Nature
  Communications}\ }\textbf {\bibinfo {volume} {5}},\ \bibinfo {pages} {5184}
  (\bibinfo {year} {2014})}\BibitemShut {NoStop}%
\bibitem [{\citenamefont {Marcus}\ \emph {et~al.}(1992)\citenamefont {Marcus},
  \citenamefont {Rimberg}, \citenamefont {Westervelt}, \citenamefont
  {Hopkins},\ and\ \citenamefont {Gossard}}]{marcus_1992}%
  \BibitemOpen
  \bibfield  {author} {\bibinfo {author} {\bibfnamefont {C.~M.}\ \bibnamefont
  {Marcus}}, \bibinfo {author} {\bibfnamefont {A.~J.}\ \bibnamefont {Rimberg}},
  \bibinfo {author} {\bibfnamefont {R.~M.}\ \bibnamefont {Westervelt}},
  \bibinfo {author} {\bibfnamefont {P.~F.}\ \bibnamefont {Hopkins}}, \ and\
  \bibinfo {author} {\bibfnamefont {A.~C.}\ \bibnamefont {Gossard}},\ }\href
  {\doibase 10.1103/PhysRevLett.69.506} {\bibfield  {journal} {\bibinfo
  {journal} {Phys. Rev. Lett.}\ }\textbf {\bibinfo {volume} {69}},\ \bibinfo
  {pages} {506} (\bibinfo {year} {1992})}\BibitemShut {NoStop}%
\bibitem [{\citenamefont {Chan}\ \emph {et~al.}(1995)\citenamefont {Chan},
  \citenamefont {Clarke}, \citenamefont {Marcus}, \citenamefont {Campman},\
  and\ \citenamefont {Gossard}}]{chan_1985}%
  \BibitemOpen
  \bibfield  {author} {\bibinfo {author} {\bibfnamefont {I.~H.}\ \bibnamefont
  {Chan}}, \bibinfo {author} {\bibfnamefont {R.~M.}\ \bibnamefont {Clarke}},
  \bibinfo {author} {\bibfnamefont {C.~M.}\ \bibnamefont {Marcus}}, \bibinfo
  {author} {\bibfnamefont {K.}~\bibnamefont {Campman}}, \ and\ \bibinfo
  {author} {\bibfnamefont {A.~C.}\ \bibnamefont {Gossard}},\ }\href {\doibase
  10.1103/PhysRevLett.74.3876} {\bibfield  {journal} {\bibinfo  {journal}
  {Phys. Rev. Lett.}\ }\textbf {\bibinfo {volume} {74}},\ \bibinfo {pages}
  {3876} (\bibinfo {year} {1995})}\BibitemShut {NoStop}%
\bibitem [{\citenamefont {Folk}\ \emph {et~al.}(1996)\citenamefont {Folk},
  \citenamefont {Patel}, \citenamefont {Godijn}, \citenamefont {Huibers},
  \citenamefont {Cronenwett}, \citenamefont {Marcus}, \citenamefont {Campman},\
  and\ \citenamefont {Gossard}}]{folk_1996}%
  \BibitemOpen
  \bibfield  {author} {\bibinfo {author} {\bibfnamefont {J.~A.}\ \bibnamefont
  {Folk}}, \bibinfo {author} {\bibfnamefont {S.~R.}\ \bibnamefont {Patel}},
  \bibinfo {author} {\bibfnamefont {S.~F.}\ \bibnamefont {Godijn}}, \bibinfo
  {author} {\bibfnamefont {A.~G.}\ \bibnamefont {Huibers}}, \bibinfo {author}
  {\bibfnamefont {S.~M.}\ \bibnamefont {Cronenwett}}, \bibinfo {author}
  {\bibfnamefont {C.~M.}\ \bibnamefont {Marcus}}, \bibinfo {author}
  {\bibfnamefont {K.}~\bibnamefont {Campman}}, \ and\ \bibinfo {author}
  {\bibfnamefont {A.~C.}\ \bibnamefont {Gossard}},\ }\href {\doibase
  10.1103/PhysRevLett.76.1699} {\bibfield  {journal} {\bibinfo  {journal}
  {Phys. Rev. Lett.}\ }\textbf {\bibinfo {volume} {76}},\ \bibinfo {pages}
  {1699} (\bibinfo {year} {1996})}\BibitemShut {NoStop}%
\bibitem [{\citenamefont {Morozov}\ \emph {et~al.}(2006)\citenamefont
  {Morozov}, \citenamefont {Novoselov}, \citenamefont {Katsnelson},
  \citenamefont {Schedin}, \citenamefont {Ponomarenko}, \citenamefont {Jiang},\
  and\ \citenamefont {Geim}}]{morozov_2006}%
  \BibitemOpen
  \bibfield  {author} {\bibinfo {author} {\bibfnamefont {S.~V.}\ \bibnamefont
  {Morozov}}, \bibinfo {author} {\bibfnamefont {K.~S.}\ \bibnamefont
  {Novoselov}}, \bibinfo {author} {\bibfnamefont {M.~I.}\ \bibnamefont
  {Katsnelson}}, \bibinfo {author} {\bibfnamefont {F.}~\bibnamefont {Schedin}},
  \bibinfo {author} {\bibfnamefont {L.~A.}\ \bibnamefont {Ponomarenko}},
  \bibinfo {author} {\bibfnamefont {D.}~\bibnamefont {Jiang}}, \ and\ \bibinfo
  {author} {\bibfnamefont {A.~K.}\ \bibnamefont {Geim}},\ }\href {\doibase
  10.1103/PhysRevLett.97.016801} {\bibfield  {journal} {\bibinfo  {journal}
  {Phys. Rev. Lett.}\ }\textbf {\bibinfo {volume} {97}},\ \bibinfo {pages}
  {016801} (\bibinfo {year} {2006})}\BibitemShut {NoStop}%
\bibitem [{\citenamefont {Albada}\ and\ \citenamefont
  {Lagendijk}(1985)}]{albada_1985}%
  \BibitemOpen
  \bibfield  {author} {\bibinfo {author} {\bibfnamefont {M.~P.~V.}\
  \bibnamefont {Albada}}\ and\ \bibinfo {author} {\bibfnamefont
  {A.}~\bibnamefont {Lagendijk}},\ }\href {\doibase
  10.1103/PhysRevLett.55.2692} {\bibfield  {journal} {\bibinfo  {journal}
  {Phys. Rev. Lett.}\ }\textbf {\bibinfo {volume} {55}},\ \bibinfo {pages}
  {2692} (\bibinfo {year} {1985})}\BibitemShut {NoStop}%
\bibitem [{\citenamefont {Wolf}\ and\ \citenamefont {Maret}(1985)}]{wolf_1985}%
  \BibitemOpen
  \bibfield  {author} {\bibinfo {author} {\bibfnamefont {P.-E.}\ \bibnamefont
  {Wolf}}\ and\ \bibinfo {author} {\bibfnamefont {G.}~\bibnamefont {Maret}},\
  }\href {\doibase 10.1103/PhysRevLett.55.2696} {\bibfield  {journal} {\bibinfo
   {journal} {Phys. Rev. Lett.}\ }\textbf {\bibinfo {volume} {55}},\ \bibinfo
  {pages} {2696} (\bibinfo {year} {1985})}\BibitemShut {NoStop}%
\bibitem [{\citenamefont {Scheffold}\ and\ \citenamefont
  {Maret}(1998)}]{scheffold_1998}%
  \BibitemOpen
  \bibfield  {author} {\bibinfo {author} {\bibfnamefont {F.}~\bibnamefont
  {Scheffold}}\ and\ \bibinfo {author} {\bibfnamefont {G.}~\bibnamefont
  {Maret}},\ }\href {\doibase 10.1103/PhysRevLett.81.5800} {\bibfield
  {journal} {\bibinfo  {journal} {Phys. Rev. Lett.}\ }\textbf {\bibinfo
  {volume} {81}},\ \bibinfo {pages} {5800} (\bibinfo {year}
  {1998})}\BibitemShut {NoStop}%
\bibitem [{\citenamefont {Schreiber}\ \emph {et~al.}(2010)\citenamefont
  {Schreiber}, \citenamefont {Cassemiro}, \citenamefont
  {Poto\ifmmode~\check{c}\else \v{c}\fi{}ek}, \citenamefont {G\'abris},
  \citenamefont {Mosley}, \citenamefont {Andersson}, \citenamefont {Jex},\ and\
  \citenamefont {Silberhorn}}]{schreiber_2010}%
  \BibitemOpen
  \bibfield  {author} {\bibinfo {author} {\bibfnamefont {A.}~\bibnamefont
  {Schreiber}}, \bibinfo {author} {\bibfnamefont {K.~N.}\ \bibnamefont
  {Cassemiro}}, \bibinfo {author} {\bibfnamefont {V.}~\bibnamefont
  {Poto\ifmmode~\check{c}\else \v{c}\fi{}ek}}, \bibinfo {author} {\bibfnamefont
  {A.}~\bibnamefont {G\'abris}}, \bibinfo {author} {\bibfnamefont {P.~J.}\
  \bibnamefont {Mosley}}, \bibinfo {author} {\bibfnamefont {E.}~\bibnamefont
  {Andersson}}, \bibinfo {author} {\bibfnamefont {I.}~\bibnamefont {Jex}}, \
  and\ \bibinfo {author} {\bibfnamefont {C.}~\bibnamefont {Silberhorn}},\
  }\href {\doibase 10.1103/PhysRevLett.104.050502} {\bibfield  {journal}
  {\bibinfo  {journal} {Phys. Rev. Lett.}\ }\textbf {\bibinfo {volume} {104}},\
  \bibinfo {pages} {050502} (\bibinfo {year} {2010})}\BibitemShut {NoStop}%
\bibitem [{\citenamefont {Ferr\'on}\ \emph {et~al.}(2017)\citenamefont
  {Ferr\'on}, \citenamefont {Dom\'{\i}nguez},\ and\ \citenamefont
  {S\'anchez}}]{ferron_2017}%
  \BibitemOpen
  \bibfield  {author} {\bibinfo {author} {\bibfnamefont {A.}~\bibnamefont
  {Ferr\'on}}, \bibinfo {author} {\bibfnamefont {D.}~\bibnamefont
  {Dom\'{\i}nguez}}, \ and\ \bibinfo {author} {\bibfnamefont {M.~J.}\
  \bibnamefont {S\'anchez}},\ }\href {\doibase 10.1103/PhysRevB.95.045412}
  {\bibfield  {journal} {\bibinfo  {journal} {Phys. Rev. B}\ }\textbf {\bibinfo
  {volume} {95}},\ \bibinfo {pages} {045412} (\bibinfo {year}
  {2017})}\BibitemShut {NoStop}%
\bibitem [{\citenamefont {Oliver}\ \emph {et~al.}(2005)\citenamefont {Oliver},
  \citenamefont {Yu}, \citenamefont {Lee}, \citenamefont {Berggren},
  \citenamefont {Levitov},\ and\ \citenamefont {Orlando}}]{Oliver2005}%
  \BibitemOpen
  \bibfield  {author} {\bibinfo {author} {\bibfnamefont {W.~D.}\ \bibnamefont
  {Oliver}}, \bibinfo {author} {\bibfnamefont {Y.}~\bibnamefont {Yu}}, \bibinfo
  {author} {\bibfnamefont {J.~C.}\ \bibnamefont {Lee}}, \bibinfo {author}
  {\bibfnamefont {K.~K.}\ \bibnamefont {Berggren}}, \bibinfo {author}
  {\bibfnamefont {L.~S.}\ \bibnamefont {Levitov}}, \ and\ \bibinfo {author}
  {\bibfnamefont {T.~P.}\ \bibnamefont {Orlando}},\ }\href {\doibase
  10.1126/science.1119678} {\bibfield  {journal} {\bibinfo  {journal}
  {Science}\ }\textbf {\bibinfo {volume} {310}},\ \bibinfo {pages} {1653}
  (\bibinfo {year} {2005})},\ \Eprint
  {http://arxiv.org/abs/http://science.sciencemag.org/content/310/5754/1653.full.pdf}
  {http://science.sciencemag.org/content/310/5754/1653.full.pdf} \BibitemShut
  {NoStop}%
\bibitem [{\citenamefont {Berns}\ \emph {et~al.}(2006)\citenamefont {Berns},
  \citenamefont {Oliver}, \citenamefont {Valenzuela}, \citenamefont {Shytov},
  \citenamefont {Berggren}, \citenamefont {Levitov},\ and\ \citenamefont
  {Orlando}}]{Berns2006}%
  \BibitemOpen
  \bibfield  {author} {\bibinfo {author} {\bibfnamefont {D.~M.}\ \bibnamefont
  {Berns}}, \bibinfo {author} {\bibfnamefont {W.~D.}\ \bibnamefont {Oliver}},
  \bibinfo {author} {\bibfnamefont {S.~O.}\ \bibnamefont {Valenzuela}},
  \bibinfo {author} {\bibfnamefont {A.~V.}\ \bibnamefont {Shytov}}, \bibinfo
  {author} {\bibfnamefont {K.~K.}\ \bibnamefont {Berggren}}, \bibinfo {author}
  {\bibfnamefont {L.~S.}\ \bibnamefont {Levitov}}, \ and\ \bibinfo {author}
  {\bibfnamefont {T.~P.}\ \bibnamefont {Orlando}},\ }\href {\doibase
  10.1103/PhysRevLett.97.150502} {\bibfield  {journal} {\bibinfo  {journal}
  {Phys. Rev. Lett.}\ }\textbf {\bibinfo {volume} {97}},\ \bibinfo {pages}
  {150502} (\bibinfo {year} {2006})}\BibitemShut {NoStop}%
\bibitem [{\citenamefont {Valenzuela}\ \emph {et~al.}(2006)\citenamefont
  {Valenzuela}, \citenamefont {Oliver}, \citenamefont {Berns}, \citenamefont
  {Berggren}, \citenamefont {Levitov},\ and\ \citenamefont
  {Orlando}}]{Valenzuela2006}%
  \BibitemOpen
  \bibfield  {author} {\bibinfo {author} {\bibfnamefont {S.~O.}\ \bibnamefont
  {Valenzuela}}, \bibinfo {author} {\bibfnamefont {W.~D.}\ \bibnamefont
  {Oliver}}, \bibinfo {author} {\bibfnamefont {D.~M.}\ \bibnamefont {Berns}},
  \bibinfo {author} {\bibfnamefont {K.~K.}\ \bibnamefont {Berggren}}, \bibinfo
  {author} {\bibfnamefont {L.~S.}\ \bibnamefont {Levitov}}, \ and\ \bibinfo
  {author} {\bibfnamefont {T.~P.}\ \bibnamefont {Orlando}},\ }\href {\doibase
  10.1126/science.1134008} {\bibfield  {journal} {\bibinfo  {journal}
  {Science}\ }\textbf {\bibinfo {volume} {314}},\ \bibinfo {pages} {1589}
  (\bibinfo {year} {2006})},\ \Eprint
  {http://arxiv.org/abs/http://science.sciencemag.org/content/314/5805/1589.full.pdf}
  {http://science.sciencemag.org/content/314/5805/1589.full.pdf} \BibitemShut
  {NoStop}%
\bibitem [{\citenamefont {Berns}\ \emph {et~al.}(2008)\citenamefont {Berns},
  \citenamefont {Rudner}, \citenamefont {Valenzuela}, \citenamefont {Berggren},
  \citenamefont {Oliver}, \citenamefont {Levitov},\ and\ \citenamefont
  {Orlando}}]{Berns2008}%
  \BibitemOpen
  \bibfield  {author} {\bibinfo {author} {\bibfnamefont {D.~M.}\ \bibnamefont
  {Berns}}, \bibinfo {author} {\bibfnamefont {M.~S.}\ \bibnamefont {Rudner}},
  \bibinfo {author} {\bibfnamefont {S.~O.}\ \bibnamefont {Valenzuela}},
  \bibinfo {author} {\bibfnamefont {K.~K.}\ \bibnamefont {Berggren}}, \bibinfo
  {author} {\bibfnamefont {W.~D.}\ \bibnamefont {Oliver}}, \bibinfo {author}
  {\bibfnamefont {L.~S.}\ \bibnamefont {Levitov}}, \ and\ \bibinfo {author}
  {\bibfnamefont {T.~P.}\ \bibnamefont {Orlando}},\ }\href
  {https://doi.org/10.1038/nature07262 http://10.0.4.14/nature07262
  https://www.nature.com/articles/nature07262{\#}supplementary-information}
  {\bibfield  {journal} {\bibinfo  {journal} {Nature}\ }\textbf {\bibinfo
  {volume} {455}},\ \bibinfo {pages} {51} (\bibinfo {year} {2008})}\BibitemShut
  {NoStop}%
\bibitem [{\citenamefont {Rudner}\ \emph {et~al.}(2008)\citenamefont {Rudner},
  \citenamefont {Shytov}, \citenamefont {Levitov}, \citenamefont {Berns},
  \citenamefont {Oliver}, \citenamefont {Valenzuela},\ and\ \citenamefont
  {Orlando}}]{Rudner2008}%
  \BibitemOpen
  \bibfield  {author} {\bibinfo {author} {\bibfnamefont {M.~S.}\ \bibnamefont
  {Rudner}}, \bibinfo {author} {\bibfnamefont {A.~V.}\ \bibnamefont {Shytov}},
  \bibinfo {author} {\bibfnamefont {L.~S.}\ \bibnamefont {Levitov}}, \bibinfo
  {author} {\bibfnamefont {D.~M.}\ \bibnamefont {Berns}}, \bibinfo {author}
  {\bibfnamefont {W.~D.}\ \bibnamefont {Oliver}}, \bibinfo {author}
  {\bibfnamefont {S.~O.}\ \bibnamefont {Valenzuela}}, \ and\ \bibinfo {author}
  {\bibfnamefont {T.~P.}\ \bibnamefont {Orlando}},\ }\href {\doibase
  10.1103/PhysRevLett.101.190502} {\bibfield  {journal} {\bibinfo  {journal}
  {Phys. Rev. Lett.}\ }\textbf {\bibinfo {volume} {101}},\ \bibinfo {pages}
  {190502} (\bibinfo {year} {2008})}\BibitemShut {NoStop}%
\bibitem [{\citenamefont {Koch}\ \emph {et~al.}(2007)\citenamefont {Koch},
  \citenamefont {Yu}, \citenamefont {Gambetta}, \citenamefont {Houck},
  \citenamefont {Schuster}, \citenamefont {Majer}, \citenamefont {Blais},
  \citenamefont {Devoret}, \citenamefont {Girvin},\ and\ \citenamefont
  {Schoelkopf}}]{koch_2007}%
  \BibitemOpen
  \bibfield  {author} {\bibinfo {author} {\bibfnamefont {J.}~\bibnamefont
  {Koch}}, \bibinfo {author} {\bibfnamefont {T.~M.}\ \bibnamefont {Yu}},
  \bibinfo {author} {\bibfnamefont {J.}~\bibnamefont {Gambetta}}, \bibinfo
  {author} {\bibfnamefont {A.~A.}\ \bibnamefont {Houck}}, \bibinfo {author}
  {\bibfnamefont {D.~I.}\ \bibnamefont {Schuster}}, \bibinfo {author}
  {\bibfnamefont {J.}~\bibnamefont {Majer}}, \bibinfo {author} {\bibfnamefont
  {A.}~\bibnamefont {Blais}}, \bibinfo {author} {\bibfnamefont {M.~H.}\
  \bibnamefont {Devoret}}, \bibinfo {author} {\bibfnamefont {S.~M.}\
  \bibnamefont {Girvin}}, \ and\ \bibinfo {author} {\bibfnamefont {R.~J.}\
  \bibnamefont {Schoelkopf}},\ }\href {\doibase 10.1103/PhysRevA.76.042319}
  {\bibfield  {journal} {\bibinfo  {journal} {Phys. Rev. A}\ }\textbf {\bibinfo
  {volume} {76}},\ \bibinfo {pages} {042319} (\bibinfo {year}
  {2007})}\BibitemShut {NoStop}%
\bibitem [{\citenamefont {Shim}\ and\ \citenamefont {Tahan}(2016)}]{shim_2016}%
  \BibitemOpen
  \bibfield  {author} {\bibinfo {author} {\bibfnamefont {Y.-P.}\ \bibnamefont
  {Shim}}\ and\ \bibinfo {author} {\bibfnamefont {C.}~\bibnamefont {Tahan}},\
  }\href {https://doi.org/10.1038/ncomms11059} {\bibfield  {journal} {\bibinfo
  {journal} {Nature Communications}\ }\textbf {\bibinfo {volume} {7}},\
  \bibinfo {pages} {11059 EP } (\bibinfo {year} {2016})}\BibitemShut {NoStop}%
\bibitem [{\citenamefont {Campbell}\ \emph {et~al.}(2019)\citenamefont
  {Campbell}, \citenamefont {Shim}, \citenamefont {Kannan}, \citenamefont
  {Winik}, \citenamefont {Kim}, \citenamefont {Yoder}, \citenamefont {Tahan},
  \citenamefont {Gustavsson},\ and\ \citenamefont {Oliver}}]{Campbell_2019}%
  \BibitemOpen
  \bibfield  {author} {\bibinfo {author} {\bibfnamefont {D.~L.}\ \bibnamefont
  {Campbell}}, \bibinfo {author} {\bibfnamefont {Y.-P.}\ \bibnamefont {Shim}},
  \bibinfo {author} {\bibfnamefont {B.}~\bibnamefont {Kannan}}, \bibinfo
  {author} {\bibfnamefont {R.}~\bibnamefont {Winik}}, \bibinfo {author}
  {\bibfnamefont {D.}~\bibnamefont {Kim}}, \bibinfo {author} {\bibfnamefont
  {J.}~\bibnamefont {Yoder}}, \bibinfo {author} {\bibfnamefont
  {C.}~\bibnamefont {Tahan}}, \bibinfo {author} {\bibfnamefont
  {S.}~\bibnamefont {Gustavsson}}, \ and\ \bibinfo {author} {\bibfnamefont
  {W.~D.}\ \bibnamefont {Oliver}},\ }\href@noop {} {\bibfield  {journal}
  {\bibinfo  {journal} {in preparation}\ } (\bibinfo {year}
  {2019})}\BibitemShut {NoStop}%
\bibitem [{\citenamefont {Barends}\ \emph {et~al.}(2013)\citenamefont
  {Barends}, \citenamefont {Kelly}, \citenamefont {Megrant}, \citenamefont
  {Sank}, \citenamefont {Jeffrey}, \citenamefont {Chen}, \citenamefont {Yin},
  \citenamefont {Chiaro}, \citenamefont {Mutus}, \citenamefont {Neill},
  \citenamefont {O'Malley}, \citenamefont {Roushan}, \citenamefont {Wenner},
  \citenamefont {White}, \citenamefont {Cleland},\ and\ \citenamefont
  {Martinis}}]{Barends2013}%
  \BibitemOpen
  \bibfield  {author} {\bibinfo {author} {\bibfnamefont {R.}~\bibnamefont
  {Barends}}, \bibinfo {author} {\bibfnamefont {J.}~\bibnamefont {Kelly}},
  \bibinfo {author} {\bibfnamefont {A.}~\bibnamefont {Megrant}}, \bibinfo
  {author} {\bibfnamefont {D.}~\bibnamefont {Sank}}, \bibinfo {author}
  {\bibfnamefont {E.}~\bibnamefont {Jeffrey}}, \bibinfo {author} {\bibfnamefont
  {Y.}~\bibnamefont {Chen}}, \bibinfo {author} {\bibfnamefont {Y.}~\bibnamefont
  {Yin}}, \bibinfo {author} {\bibfnamefont {B.}~\bibnamefont {Chiaro}},
  \bibinfo {author} {\bibfnamefont {J.}~\bibnamefont {Mutus}}, \bibinfo
  {author} {\bibfnamefont {C.}~\bibnamefont {Neill}}, \bibinfo {author}
  {\bibfnamefont {P.}~\bibnamefont {O'Malley}}, \bibinfo {author}
  {\bibfnamefont {P.}~\bibnamefont {Roushan}}, \bibinfo {author} {\bibfnamefont
  {J.}~\bibnamefont {Wenner}}, \bibinfo {author} {\bibfnamefont {T.~C.}\
  \bibnamefont {White}}, \bibinfo {author} {\bibfnamefont {A.~N.}\ \bibnamefont
  {Cleland}}, \ and\ \bibinfo {author} {\bibfnamefont {J.~M.}\ \bibnamefont
  {Martinis}},\ }\href {\doibase 10.1103/PhysRevLett.111.080502} {\bibfield
  {journal} {\bibinfo  {journal} {Phys. Rev. Lett.}\ }\textbf {\bibinfo
  {volume} {111}},\ \bibinfo {pages} {080502} (\bibinfo {year}
  {2013})}\BibitemShut {NoStop}%
\bibitem [{\citenamefont {Hutchings}\ \emph {et~al.}(2017)\citenamefont
  {Hutchings}, \citenamefont {Hertzberg}, \citenamefont {Liu}, \citenamefont
  {Bronn}, \citenamefont {Keefe}, \citenamefont {Brink}, \citenamefont {Chow},\
  and\ \citenamefont {Plourde}}]{Hutchings2017}%
  \BibitemOpen
  \bibfield  {author} {\bibinfo {author} {\bibfnamefont {M.~D.}\ \bibnamefont
  {Hutchings}}, \bibinfo {author} {\bibfnamefont {J.~B.}\ \bibnamefont
  {Hertzberg}}, \bibinfo {author} {\bibfnamefont {Y.}~\bibnamefont {Liu}},
  \bibinfo {author} {\bibfnamefont {N.~T.}\ \bibnamefont {Bronn}}, \bibinfo
  {author} {\bibfnamefont {G.~A.}\ \bibnamefont {Keefe}}, \bibinfo {author}
  {\bibfnamefont {M.}~\bibnamefont {Brink}}, \bibinfo {author} {\bibfnamefont
  {J.~M.}\ \bibnamefont {Chow}}, \ and\ \bibinfo {author} {\bibfnamefont
  {B.~L.~T.}\ \bibnamefont {Plourde}},\ }\href {\doibase
  10.1103/PhysRevApplied.8.044003} {\bibfield  {journal} {\bibinfo  {journal}
  {Phys. Rev. Applied}\ }\textbf {\bibinfo {volume} {8}},\ \bibinfo {pages}
  {044003} (\bibinfo {year} {2017})}\BibitemShut {NoStop}%
\bibitem [{\citenamefont {Orlando}\ \emph {et~al.}(1999)\citenamefont
  {Orlando}, \citenamefont {Mooij}, \citenamefont {Tian}, \citenamefont
  {van~der Wal}, \citenamefont {Levitov}, \citenamefont {Lloyd},\ and\
  \citenamefont {Mazo}}]{Orlando1999}%
  \BibitemOpen
  \bibfield  {author} {\bibinfo {author} {\bibfnamefont {T.~P.}\ \bibnamefont
  {Orlando}}, \bibinfo {author} {\bibfnamefont {J.~E.}\ \bibnamefont {Mooij}},
  \bibinfo {author} {\bibfnamefont {L.}~\bibnamefont {Tian}}, \bibinfo {author}
  {\bibfnamefont {C.~H.}\ \bibnamefont {van~der Wal}}, \bibinfo {author}
  {\bibfnamefont {L.~S.}\ \bibnamefont {Levitov}}, \bibinfo {author}
  {\bibfnamefont {S.}~\bibnamefont {Lloyd}}, \ and\ \bibinfo {author}
  {\bibfnamefont {J.~J.}\ \bibnamefont {Mazo}},\ }\href {\doibase
  10.1103/PhysRevB.60.15398} {\bibfield  {journal} {\bibinfo  {journal} {Phys.
  Rev. B}\ }\textbf {\bibinfo {volume} {60}},\ \bibinfo {pages} {15398}
  (\bibinfo {year} {1999})}\BibitemShut {NoStop}%
\bibitem [{sup()}]{supp-mat}%
  \BibitemOpen
  \href@noop {} {\bibinfo  {journal} {see supplementary material.}\
  }\BibitemShut {NoStop}%
\bibitem [{\citenamefont {Chiorsescu}\ \emph {et~al.}(2003)\citenamefont
  {Chiorsescu}, \citenamefont {Nakamura}, \citenamefont {Harmans},\ and\
  \citenamefont {Mooij}}]{Chiorescu2003}%
  \BibitemOpen
\bibfield  {journal} {  }\bibfield  {author} {\bibinfo {author} {\bibfnamefont
  {I.}~\bibnamefont {Chiorsescu}}, \bibinfo {author} {\bibfnamefont
  {Y.}~\bibnamefont {Nakamura}}, \bibinfo {author} {\bibfnamefont {C.~J.
  P.~M.}\ \bibnamefont {Harmans}}, \ and\ \bibinfo {author} {\bibfnamefont
  {J.~E.}\ \bibnamefont {Mooij}},\ }\href@noop {} {\bibfield  {journal}
  {\bibinfo  {journal} {Science}\ }\textbf {\bibinfo {volume} {299}},\ \bibinfo
  {pages} {1869} (\bibinfo {year} {2003})}\BibitemShut {NoStop}%
\bibitem [{\citenamefont {Bylander}\ \emph {et~al.}(2011)\citenamefont
  {Bylander}, \citenamefont {Gustavsson}, \citenamefont {Yan}, \citenamefont
  {Yoshihara}, \citenamefont {Harrabi}, \citenamefont {Fitch}, \citenamefont
  {Cory}, \citenamefont {Nakamura}, \citenamefont {Tsai},\ and\ \citenamefont
  {Oliver}}]{Bylander2011}%
  \BibitemOpen
  \bibfield  {author} {\bibinfo {author} {\bibfnamefont {J.}~\bibnamefont
  {Bylander}}, \bibinfo {author} {\bibfnamefont {S.}~\bibnamefont
  {Gustavsson}}, \bibinfo {author} {\bibfnamefont {F.}~\bibnamefont {Yan}},
  \bibinfo {author} {\bibfnamefont {F.}~\bibnamefont {Yoshihara}}, \bibinfo
  {author} {\bibfnamefont {K.}~\bibnamefont {Harrabi}}, \bibinfo {author}
  {\bibfnamefont {G.}~\bibnamefont {Fitch}}, \bibinfo {author} {\bibfnamefont
  {D.~G.}\ \bibnamefont {Cory}}, \bibinfo {author} {\bibfnamefont
  {Y.}~\bibnamefont {Nakamura}}, \bibinfo {author} {\bibfnamefont {J.-S.}\
  \bibnamefont {Tsai}}, \ and\ \bibinfo {author} {\bibfnamefont {W.~D.}\
  \bibnamefont {Oliver}},\ }\href {\doibase 10.1038/nphys1994} {\bibfield
  {journal} {\bibinfo  {journal} {Nature Physics}\ }\textbf {\bibinfo {volume}
  {7}},\ \bibinfo {pages} {565} (\bibinfo {year} {2011})}\BibitemShut {NoStop}%
\bibitem [{\citenamefont {Shevchenko}\ \emph {et~al.}(2010)\citenamefont
  {Shevchenko}, \citenamefont {Ashhab},\ and\ \citenamefont
  {Nori}}]{shevchenko_2010}%
  \BibitemOpen
  \bibfield  {author} {\bibinfo {author} {\bibfnamefont {S.}~\bibnamefont
  {Shevchenko}}, \bibinfo {author} {\bibfnamefont {S.}~\bibnamefont {Ashhab}},
  \ and\ \bibinfo {author} {\bibfnamefont {F.}~\bibnamefont {Nori}},\ }\href
  {\doibase https://doi.org/10.1016/j.physrep.2010.03.002} {\bibfield
  {journal} {\bibinfo  {journal} {Physics Reports}\ }\textbf {\bibinfo {volume}
  {492}},\ \bibinfo {pages} {1 } (\bibinfo {year} {2010})}\BibitemShut
  {NoStop}%
\bibitem [{\citenamefont {Baranger}\ \emph {et~al.}(1993)\citenamefont
  {Baranger}, \citenamefont {Jalabert},\ and\ \citenamefont
  {Stone}}]{Baranger_1993}%
  \BibitemOpen
  \bibfield  {author} {\bibinfo {author} {\bibfnamefont {H.~U.}\ \bibnamefont
  {Baranger}}, \bibinfo {author} {\bibfnamefont {R.~A.}\ \bibnamefont
  {Jalabert}}, \ and\ \bibinfo {author} {\bibfnamefont {A.~D.}\ \bibnamefont
  {Stone}},\ }\href {\doibase 10.1103/PhysRevLett.70.3876} {\bibfield
  {journal} {\bibinfo  {journal} {Phys. Rev. Lett.}\ }\textbf {\bibinfo
  {volume} {70}},\ \bibinfo {pages} {3876} (\bibinfo {year}
  {1993})}\BibitemShut {NoStop}%
\end{thebibliography}%


\begin{thebibliography}{0}
\bibitem{1}  A. L. Gramajo, D. Campbell, B. Kannan, D. K. Kim, A. Melville, B. Niedzielski, J. L. Yoder, M. J. S\'anchez, D. Dom\'inguez, S. Gustavsson and W. D. Oliver (2019).
\bibitem{hutchings} M. D. Hutchings, J. B. Hertzberg, Y. Liu,1 N. T. Bronn, G. A. Keefe, Markus Brink, Jerry M. Chow, and B. L. T. Plourde, Phys. Rev. App. 8, 044003 (2017).
\bibitem{propagator} W. Wustmann, Doctor of Philosophy thesis: Statistical mechanics of time-periodic quantum systems, Technische Universit\"at Dresden (2010).
\bibitem{2}  S. Shevchenko, S. Ashhab, and F. Nori, Phys. Rep. 492, 1 (2010).
\bibitem{3}  D. M. Berns, Doctor of Philosophy thesis: Large amplitude driving of a persistent current qubit, Massachusetts Institute of Technology (2008).
\bibitem{4}  A. Ferr\'on, D. Dom\'inguez and M. J. S\'anchez , Phys. Rev. B 95, 045412 (2017).
\bibitem{9} C. W. J. Beenakker and J. A. Melsen, Phys. Rev. B 50 (1994).
\bibitem{5} Harold U. Baranger, Rodolfo A. Jalabert, and A. Douglas Stone. Phys. Rev. Lett. 70, 3876 ( 1993).
\bibitem{6} C. W. J. Beenakker, Rev. Mod. Phys. 69, 731  (1997).
\bibitem{7}  S. Datta, Electronic Transport in Mesoscopic Systems, Cambridge Studies in Semiconductor Physics and Microelectronic Engineering (Cambridge University Press, 1995).
\bibitem{8} D. Ferry and S. M. Goodnick, Transport in Nanostructures, Cambridge Studies in Semiconductor Physics and Microelectronic Engineering (Cambridge University Press, 1997).


\end{thebibliography}


\pagebreak
\clearpage
\newpage
\widetext

\begin{center}
\textbf{\large Supplementary information: Quantum emulation of coherent backscattering in a system of superconducting qubits}
\end{center}
\setcounter{equation}{0}
\setcounter{figure}{0}
\setcounter{table}{0}
\setcounter{page}{1}
\makeatletter
\renewcommand{\theequation}{S\arabic{equation}}
\renewcommand{\thefigure}{S\arabic{figure}}
\renewcommand{\bibnumfmt}[1]{[S#1]}
\renewcommand{\citenumfont}[1]{S#1}

\section{\label{sec:A} Sinusoidal approximation for the diabatic frequency $\varepsilon$}
\setcounter{figure}{0}

In this section, we derive the sinusoidal expression
\begin{equation}
    \varepsilon(t) \approx  \delta \omega \sin \left[ 2 \pi f_{\textrm{dc}}(t) \right] + {\omega},
    \label{eq:A1}
\end{equation}
where $\delta \omega = ({\omega}^{\textrm{max}} - {\omega}^{\textrm{min}})/2$ and ${\omega} = ({\omega}^{\textrm{max}} + {\omega}^{\textrm{min}}$)/2, with ${\omega}^{\textrm{max/min}}$ the average of $\omega_{\textrm{a}}^{\textrm{max/min}}$ and $\omega_{\textrm{b}}^{\textrm{max/min}}$, see Ref.\cite{1} for further details.

To obtain Eq.\eqref{eq:A1}, we start by considering the general expression \cite{hutchings} for the diabatic frequency of one of the transmons, say $Q_{\textrm{a}}$,
\begin{equation}
	\varepsilon_{\textrm{a}}(t) =  \sqrt{E_{J\Sigma,\textrm{a}} \cos(\pi f_{\textrm{dc}}(t) ) \sqrt{1+d^{2}\tan(\pi f_{\textrm{dc}}(t) )^{2}}},
\label{eq:A2}
\end{equation}
where $E_{J\Sigma,\textrm{a}} = \sqrt{E_{J1,\textrm{a}} + E_{J2,\textrm{a}}}$, with  $E_{J1,\textrm{a}},E_{J2,\textrm{a}}$ the Josephson energies of each junction, satisfying $E_{J1,\textrm{a}} = \alpha E_{J2,\textrm{a}}$ and $E_{J1,\textrm{a}}\gg E_{J2,\textrm{a}}$. The parameter $d$ follows thus the relation
\begin{equation}
    \begin{aligned}
        d= \frac{\alpha - 1}{\alpha + 1}.
    \end{aligned}	
	\label{eq:A1_a}
\end{equation}
Furthermore, since we are working in the limit of the large area ratio of $\alpha$ of the junctions, then $d\rightarrow 1$. Notice that the two transmons are well-matched, with maximum frequencies
$\omega_{\textrm{a}}^{\textrm{max}}/2\pi$ = 3.8250 GHz and $\omega_{\textrm{b}}^{\textrm{max}}/2\pi$ = 3.8218 GHz and minimum frequencies
$\omega_{\textrm{a}}^{\textrm{min}}/2\pi$ = 3.5401 GHz and $\omega_{\textrm{b}}^{\textrm{min}}/2\pi$ = 3.5365 GHz. Therefore, for the moment, we only focus on one transmon.

As follows, we first start with the expression:
\begin{equation}
    \begin{aligned}
     E_{J \Sigma, \textrm{a}} \cos(x) \sqrt{1+d^2 \tan^2(x)},
    \end{aligned}	
	\label{eq:A3}
\end{equation}
corresponding to the function inside the square root of Eq.\eqref{eq:A2} and where we define $x=\pi f_{\textrm{dc}} (t)$ to simplify notation. Taking the limit $d \rightarrow 1$, the previous equation remains
\begin{equation}
    \begin{aligned}
         \lim_{d\to1} \sqrt{1+d^2 \tan^2(x)} \rightarrow \sqrt{\sec(x)^{2}} + (d-1) \frac{\tan(x)^{2} }{\sqrt{\sec(x)^{2}}} + O((d-1)^{2}).
    \end{aligned}	
	\label{eq:A4}
\end{equation}
Notice that to obtain the r.h.s. term of Eq.\eqref{eq:A4}, we performed an Taylor series expansion around $d=1$, which means
\begin{equation}
    \begin{aligned}
         f(d,x) = \sqrt{1+d^2 \tan^2(x)} = \sum_{n=0}^{\infty} \frac{\partial^n f(d,x)}{\partial^n d} \frac{(d-1)^{n}}{n!},
    \end{aligned}	
	\label{eq:A5}
\end{equation}
$n\in\mathbb{N}$, and we only keep the first two terms.

Replacing \eqref{eq:A4} into \eqref{eq:A3}, we get
\begin{equation}
\begin{aligned}
   E_{J\Sigma,\textrm{a}} \cos(x) \left( \sec(x) + (d-1) \frac{\tan(x)^{2} }{\sec(x)} \right) \rightarrow  E_{J\Sigma,\textrm{a}} +  E_{J\Sigma,\textrm{a}} (d-1) \sin^{2}(x),
  \end{aligned}	
	\label{eq:A6}
\end{equation}

Using the previous result, the Eq.\eqref{eq:A2} transforms as
\begin{equation}
\begin{aligned}
   \varepsilon(t) \approx \sqrt{E_{J\Sigma,\textrm{a}} +  E_{J\Sigma,\textrm{a}} (d-1) \sin^{2}(x)}.
  \end{aligned}	
	\label{eq:A7}
\end{equation}

Going a step further and using the Taylor series expansion
\begin{equation}
\begin{aligned}
    \lim_{d\to1} \sqrt{b + (d-1) a} \rightarrow \sqrt{b} + \frac{a}{2\sqrt{b}} (d-1) + O((d-1)^{2}),
  \end{aligned}	
	\label{eq:A8}
\end{equation}  where $b=E_{J\Sigma,\textrm{a}}$ and $a=E_{J\Sigma,\textrm{a}}\sin(x)^{2}$, the Eq.\eqref{eq:A7} can be written as
\begin{equation}
\begin{aligned}
   \varepsilon_{\textrm{a}}(t) \approx  \sqrt{E_{J\Sigma,\textrm{a}}} +  \frac{\sqrt{E_{J\Sigma,\textrm{a}}} (d-1)}{2} \sin^{2}(x).
  \end{aligned}	
	\label{eq:A9}
\end{equation} Further using the trigonometric relation $\sin^{2}(x) = \left( 1- \cos(2x) \right)/2$,
\begin{equation}
\begin{aligned}
  \varepsilon_{\textrm{a}}(t) \approx \sqrt{E_{J\Sigma,\textrm{a}}}\left( 1 + \frac{d-1}{4}\right) -  \frac{\sqrt{E_{J\Sigma,\textrm{a}}} (d-1)}{4} \cos(2x).
  \end{aligned}	
	\label{eq:A10}
\end{equation}  Since, we define $x=\pi f_{\textrm{dc}}$,
\begin{equation}
\begin{aligned}
   \varepsilon_{\textrm{a}}(t) &\approx  \sqrt{E_{J\Sigma,\textrm{a}}}\left( 1 + \frac{d-1}{4}\right) -  \frac{\sqrt{E_{J\Sigma,\textrm{a}}} (d-1)}{4} \cos \left[ 2\pi f_{\textrm{dc}}(t) \right],\\
   &\approx  \sqrt{E_{J\Sigma,\textrm{a}}}\left( 1 - \frac{1-d}{4}\right) +  \frac{\sqrt{E_{J\Sigma,\textrm{a}}} (1-d)}{4}\cos \left[ 2\pi f_{\textrm{dc}}(t) \right].
  \end{aligned}	
	\label{eq:A11}
\end{equation}

Now, we proceed to demonstrate the following relations:
\begin{equation}
\begin{aligned}
    \sqrt{E_{J\Sigma,\textrm{a}}}\left( 1 - \frac{1-d}{4}\right) &\approx \frac{{\omega}^{\textrm{max}}+{\omega}^{\textrm{min}}}{2},\\
     \frac{\sqrt{E_{J\Sigma,\textrm{a}}} (1-d)}{4} &\approx \frac{{\omega}^{\textrm{max}}-{\omega}^{\textrm{min}}}{2}.
  \end{aligned}	
	\label{eq:A12}
\end{equation} For such a purpose, we start by considering the definitions:
\begin{equation}
\begin{aligned}
   {\omega}^{\textrm{max}}_{\textrm{a}}&=\sqrt{E_{J1,\textrm{a}} + E_{J2,\textrm{a}}} =  \sqrt{E_{J\Sigma,\textrm{a}}},\\
   {\omega}^{\textrm{min}}_{\textrm{a}}&=\sqrt{E_{J1,\textrm{a}} - E_{J2,\textrm{a}}}.\\
  \end{aligned}	
	\label{eq:A13}
\end{equation} Remember that $E_{J1,\textrm{a}}\gg E_{J2,\textrm{a}}$, with $\alpha=E_{J1,\textrm{a}}/E_{J2,\textrm{a}}$. Moreover, using Eq.\eqref{eq:A13} into Eq.\eqref{eq:A1_a}, we obtain the contraint
\begin{equation}
\begin{aligned}
    \sqrt{d}= \frac{\omega^{\textrm{max}}_{\textrm{a}}}{\omega^{\textrm{min}}_{\textrm{a}}}.
      \end{aligned}	
	\label{eq:A15}
\end{equation}

Thus, replacing \eqref{eq:A15}   into the r.h.s. of \eqref{eq:A12}, we get
\begin{equation}
\begin{aligned}
    \frac{  {\omega}^{\textrm{max}}_{\textrm{a}} +   {\omega}^{\textrm{min}}_{\textrm{a}}  }{2} &= \omega_{max}\frac{1+\sqrt{d}}{2} = \sqrt{E_{J\Sigma,\textrm{a}}}\frac{1+\sqrt{d}}{2} ,\\
    \frac{   {\omega}^{\textrm{max}}_{\textrm{a}}  -    {\omega}^{\textrm{min}}_{\textrm{a}} }{2} &= \omega_{max}\frac{1-\sqrt{d}}{2} = \sqrt{E_{J\Sigma,\textrm{a}}}\frac{1-\sqrt{d}}{2}.
  \end{aligned}	
	\label{eq:A16}
\end{equation}  This last expression is quite similar to Eq.\eqref{eq:A9}, but we need to work a little more to obtain the same equality.

Under the limit $d\rightarrow 1$, we can perform the following approach
\begin{equation}
\begin{aligned}
     \lim_{d\to1} \sqrt{d} \rightarrow 1 + \frac{d-1}{2} + O((d-1)^{2}).
  \end{aligned}	
	\label{eq:A17}
\end{equation} Replacing this expression into \eqref{eq:A16}, we obtain
\begin{equation}
\begin{aligned}
    \frac{  {\omega}^{\textrm{max}}_{\textrm{a}} +  {\omega}^{\textrm{min}}_{\textrm{a}}  }{2} &\approx \sqrt{E_{J\Sigma,\textrm{a}}}\left( 1 - \frac{1-d}{4}\right) ,\\
    \frac{  {\omega}^{\textrm{max}}_{\textrm{a}}  -  {\omega}^{\textrm{min}}_{\textrm{a}}   }{2} &\approx \sqrt{E_{J\Sigma,\textrm{a}}}\frac{1-d}{4}.
  \end{aligned}	
	\label{eq:A18}
\end{equation}  Then, the Eq.\eqref{eq:A12} is satisfied.

We finally reach to the diabatic frequency for the transmon $Q_{\textrm{a}}$
\begin{equation}
\begin{aligned}
   \varepsilon_{\textrm{a}}(t)  &\approx   \frac{  {\omega}^{\textrm{max}}_{\textrm{a}} -   {\omega}^{\textrm{min}}_{\textrm{a}}  }{2}\cos \left[ 2\pi f_{\textrm{dc}}(t) \right] +\frac{  {\omega}^{\textrm{max}}_{\textrm{a}} +   {\omega}^{\textrm{min}}_{\textrm{a}}   }{2},\\
  \end{aligned}	
	\label{eq:A19}
\end{equation}  Following the same procedure, the diabatic frequency for the transmon $Q_{\textrm{b}}$ remains
\begin{equation}
\begin{aligned}
   \varepsilon_{\textrm{b}}(t)  &\approx   \frac{  {\omega}^{\textrm{max}}_{\textrm{b}} -   {\omega}^{\textrm{min}}_{\textrm{b}}  }{2}\cos \left[ 2\pi f_{\textrm{dc}}(t) \right] +\frac{  {\omega}^{\textrm{max}}_{\textrm{b}} +   {\omega}^{\textrm{min}}_{\textrm{b}}   }{2},\\
  \end{aligned}	
	\label{eq:A20}
\end{equation}

Furthermore, we choose to work with the average diabatic frequency $\varepsilon(t) = \left( \varepsilon_{\textrm{a}}(t) + \varepsilon_{\textrm{b}}(t)\right) /2 $, since the transmons frequencies match quite well $\omega_{\textrm{a}}^{\textrm{max/min}}\approx \omega_{\textrm{b}}^\textrm{max/min}$. Finally, we obtain
\begin{equation}
\begin{aligned}
   \varepsilon(t)  &\approx   \frac{  \bar{\omega}^{\textrm{max}} -   \bar{\omega}^{\textrm{min}}  }{2}\cos \left[ 2\pi f_{\textrm{dc}}(t) \right] + \frac{  \bar{\omega}^{\textrm{max}} +   \bar{\omega}^{\textrm{min}}   }{2},\\
    &\approx   \delta \omega \cos \left[ 2\pi f_{\textrm{dc}}(t) \right]  + \overline{\omega}.
  \end{aligned}	
	\label{eq:A21}
\end{equation}

\textit{ Note 1:} The numerical results presented in our work \cite{1} has been performed using the propagator method \cite{propagator}.  Since the effective Hamiltonian is periodic in time, we can further use the Floquet formalism \cite{2} to solve the system time-evolution.

As follows, and without going into technical details, the system dynamics can be compute solving the time-evolution of the propagator $U(t+\tau,t)$, with $\tau$ the period of the system Hamiltonian. If the Hamiltonian can be split as $\hat{H}_{\text{eff}}(t) = \hat{H}_{0} + \hat{V}(t)$, then $U(t+\tau,t)$ can be factorized into a kinetic and a potential part. Thus, we reach to the expansion
\begin{equation}
\begin{aligned}
	U(t+\delta t,t)  &=  \text{e}^{-i \hat{H}_{0}\frac{\delta t}{2}} \text{e}^{-i \hat{V}(t+ \frac{\delta t}{2}){\delta t}} \text{e}^{-i \hat{H}_{0}\frac{\delta t}{2}},\\
	 &= \Pi_{n=0}^{N-1} U\left( (n+1) \delta t, n\delta t\right)
\end{aligned}	
 \label{eq:U_exp}
\end{equation} with $\delta t= \tau/N$ the time interval.

Such factorization carries on several difficulties when considering the full expression of $\varepsilon(t)$ \eqref{eq:A2}, which is the reason why we employ the approximate equation \eqref{eq:A1}.

\vspace{0.5cm}
\textit{ Note 2:} Furthermore, to obtain such numerical and analytical results, we set for convenience the expression
\begin{equation}
\begin{aligned}
   \varepsilon(t)  &\approx  \delta \omega \sin \left[ 2\pi f_{\textrm{dc}}(t) \right],
  \end{aligned}	
	\label{eq:A22}
\end{equation} where the constant frequency $\overline{\omega}$ is absorbed, and we turn $\cos \left[ 2\pi f_{\textrm{dc}}(t) \right] \rightarrow \sin \left[ 2\pi f_{\textrm{dc}}(t) \right]$. By doing this, we are setting the locations of the avoided crossing $\Delta$ at $n\Phi_{0}$, with $ n \in \mathbb{Z} $. Notice that  $\cos \left[ 2\pi f_{\textrm{dc}}(t) \right]$ the avoided crossings are located at $n\Phi_{0}/2$. The corresponding experimental value of $\overline{\omega}$ is $\bar{\omega}/2\pi = 3.6809$ GHz.

\section{\label{sec:B} Landau-Zener-St\"uckelberg Interferometry with single driving: slow passage limit}

In this section, we briefly analyze the system dynamics of our encoded qubit driven by a single microwave field. In this way, we present experimental and numerical results along with a short analytical description.

\begin{figure}[hbt!]
	\centering
	\includegraphics[width=17cm]{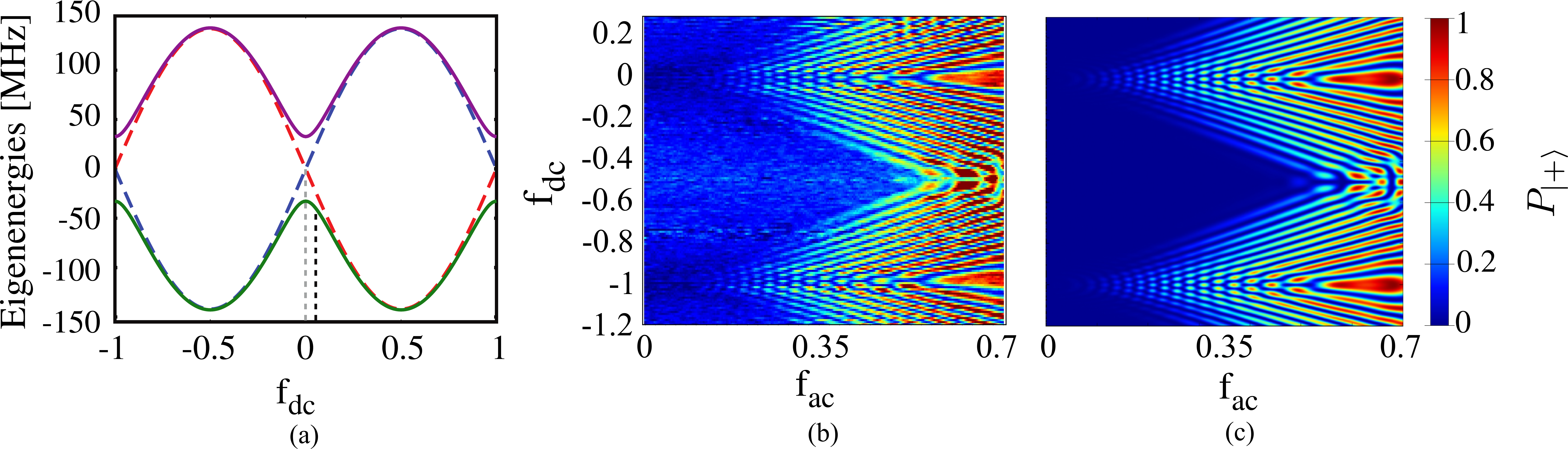}
	\caption{\textbf{Resonance pattern of transition probability}. \textbf{(a)} Transition probability measurement $P_{|+\rangle}(t_{exp})$ in terms of $f_{\textrm{dc}}$ and $f_{\textrm{ac}}$. \textbf{(b)} Numerical results. \textbf{(c)} Plot of the energy-levels in an extended range of $f_{\textrm{dc}}$. When the qubit is tune near to an avoided crossing and it is weakly driving, the  resonance pattern obeys the dynamics presented in Eq.(\ref{eq:B2}). Only when the qubit is tuned near to $f_{\textrm{dc}}=0.5\Phi_0$ and it is strongly driving ($f_{\textrm{dc}}> 0.35 \Phi_{0}$), the transition probability in the space $\left[ f_{\textrm{ac}} , f_{\textrm{dc}}\right]$ manifests another kind of resonance pattern due to the presence of a second avoided crossing.}
	\label{fig:A1}
\end{figure}

As it was presented in  \cite{1}, the system of work can be modeled as an effective Two-Level System (TLS) driven by an external periodic signal described by the effective Hamiltonian
\begin{equation}
\begin{aligned}
 \hat{H}_{\text{eff}}(t)/\hbar&= -\frac{\varepsilon(t)}{2} \hat{\sigma}_{z} - \frac{\Delta}{2},
 \label{eq:B1}
 \end{aligned}
\end{equation} with $\varepsilon(t)$ defined in Eq.\eqref{eq:A22}, $f(t)= f_{\textrm{dc}} + f_{\textrm{ac}}(t)$ and $f_{\textrm{ac}}(t)$ as the external driving. While $f_{\textrm{ac}}(t)$ can take any form, in this section we only consider the case of single driving $f_{\textrm{ac}}(t) = f_{\textrm{ac}}\cos(\omega t)$. Notice that the Hamiltonian \eqref{eq:B1} is written in the manifold basis $\{ |g_{\textrm{a}}e_{\textrm{b}}\rangle, |e_{\textrm{a}}g_{\textrm{b}}\rangle \}$. The corresponding energy spectrum as a function of $f_{\textrm{dc}}$ is plotted in Fig.~\ref{fig:A1}a. The spectrum displays a periodic behaviour in terms of $f_{\textrm{dc}}$, where the avoided crossings are periodically located at $f_{\textrm{dc}}= n \Phi_{0}$, $ n \in \mathbb{Z} $.

Fig.~\ref{fig:A1}b shows the measurement of the LZS interferometric pattern \cite{2}, plotting the transition probability  $P_{|+\rangle} (t_{exp})$  as function of $f_{\textrm{dc}}$ and $f_{\textrm{ac}}$, after one period of the driving $t_{exp} = 2\pi/\omega$.  The system has been initially prepared in the ground state $|-\rangle$ for each $f_{\textrm{dc}}$ value. By its side, it is presented the corresponding  numerical simulation results in the  solution of  the Schr\"odinger equation. As shown,  Fig.~\ref{fig:A1}c reproduces the experimental results quite well.

The calculation of an analytic expression for the time-evolution of transition probability carries on several difficulties. In this way, it is helpful to make use of different theoretical approaches to describe the dynamics.  Since the driving frequency, $\omega/2\pi \approx 10$MHz is smaller than the minimum energy splitting $\Delta/2\pi \approx 65,4 $Mhz, the most suitable approach corresponds to the slow limit passage \cite{2}. In this regime, the interference fringes in the transition probability satisfy the resonance condition
\begin{equation}
	\xi_{1} + \xi_{2}= k\pi,
 \label{eq:B2}
\end{equation} $\forall k \in \mathbb{Z}$, with $\xi_{1} = \int_{t_{1}}^{t_{2}} \Omega(t) dt$ and $\xi_{2} = \int_{t_{2}}^{t_{1} + 2\pi/\omega} \Omega(t) dt$, $\Omega(t) = \sqrt{\varepsilon(t)^{2} + \Delta^{2}}/2$. Notice that the integrals $\xi_{1}$ and $\xi_{2}$ cannot be easily evaluated. However, the resonance condition describes arcs around the point $\left(f_{\textrm{ac}},f_{\textrm{dc}}\right) = \left(0,n\Phi_{0} \right)$, observed in Fig.\ref{fig:A1}b and Fig.\ref{fig:A1}c when the qubit is driven quite close to each avoided crossing, see Fig.~\ref{fig:A1}c to identify the eigenenergy spectrum.

It should be noticed that, for our case, the picture described above \cite{2} is only valid for $f_{\textrm{ac}} \ll 0.5 \Phi_{0}$, when the qubit is driving near to each avoided crossing. Beyond the upper-bound $f_{\textrm{ac}} \gg 0.5 \Phi_{0}$, the interference pattern as a function of $f_{\textrm{dc}}$ and $f_{\textrm{ac}}$ notably changes, since the geometry of the system becomes relevant in the system dynamics. These effects can be observed in Figs.~\ref{fig:A1}b and ~\ref{fig:A1}c when the qubit is driven too far away from $f_{\textrm{dc}}=0$, reaching the second avoided crossing, see Fig.~\ref{fig:A1}a.



\section{\label{sec:C} Breaking time-reversal symmetry: analytical calculation of transition rate}

\hspace{0.6cm}Similarly to previous results \cite{3,4}, one can approximately calculate the transition rate from the ground state to the excited state via perturbation theory.

We are going to consider the system Hamiltonian presented in the section above, see Eq.(\ref{eq:B1}),   with the biharmonic driving  $f_{\textrm{ac}}(t)=f_{\textrm{ac}1}\cos(\omega_{1} t) +f_{ac2} \cos(\omega_{2} t + \alpha)$, where $\omega_{1}=\omega$ and $\omega_{2}=2\omega$. In order to simplify the calculations, we can work  under the assumption that the qubit is only driven near to one of the avoided crossings, whereby the Hamiltonian (\ref{eq:B1}) becomes linear around $f_{\textrm{dc}} = n\Phi_{0}$. In this way, we can work with the simplifying Hamiltonian, for $n=0$,
\begin{equation}
\begin{aligned}
 \hat{H'}_{\text{eff}}(t)/\hbar&\approx -\frac{2 \pi \delta \omega  f(t)/\Phi_{0}}{2} \hat{\sigma}_{z} - \frac{\Delta}{2} \hat{\sigma}_{x},\\
 &\approx  -\frac{h(t)}{2} \hat{\sigma}_{z} - \frac{\Delta}{2} \hat{\sigma}_{x},
\end{aligned}
 \label{eq:C1}
\end{equation} with
$$h(t) = 2 \pi \delta \omega  f(t) = \varepsilon_{0} + g(t),$$
 defining
 $$\varepsilon_{0}= 2\pi \delta \omega  f_{\textrm{dc}}$$
  and
  $$g(t) = 2\pi \delta \omega  \tilde{f}(t) = A_{1}\cos(\omega_{1} t) + A_{2}\cos(\omega_{2} t + \alpha)$$  with $A_{1}= 2\pi \delta \omega  f_{\textrm{ac}1}$ and $A_{2}=2  \pi \delta \omega  f_{\textrm{ac}2}$.

Applying the unitary transformation $\hat{R}=e^{-i\phi(t) \sigma^{(i)}_{z}/2}$, $\phi(t)=\int_{0}^{t} h(t) \,dt$,  to the linearized Hamiltonian (\ref{eq:C1}), we obtain
\begin{equation}
 \hat{\mathcal{H}}(t)= -\frac{\Delta(t)}{2} \hat{\sigma}_{+} - \frac{\Delta(t)^{*}}{2} \hat{\sigma}_{-},
 \label{eq:C2}
\end{equation} with $\Delta(t)=\Delta e ^{-i \phi(t)}$. Notice that this procedure has brought the problem to the interaction picture corresponding to a rotation of the Hamiltonian into a rotating framework.

For another part,  we can define the  transition rate between the ground $|-\rangle$ and the excited $|+\rangle$ states (diabatic basis) as
\begin{equation}
 W = \frac{d P_{ |+\rangle } (t)}{dt}=  \frac{d P_{|-\rangle \rightarrow |+\rangle } (t)}{dt} ,
 \label{eq:C3}
\end{equation} with $P_{|-\rangle \rightarrow |+\rangle}= \left|\langle-| U_{I}(t,0)  |+\rangle \right |^{2}$ the transition probability. In this way, W remains
\begin{equation}
 W = \frac{d}{dt}  \left| \langle-| U_{I}(t,0)  |+\rangle \right|^{2} .
 \label{eq:C4}
\end{equation}

Under the assumption of  $\Delta \rightarrow 0$, the evolution operator can be expanded to first order in $\Delta$ \cite{3,4}, thus obtaining
\begin{equation}
 U_{I}(t,0) = 1 - i\int_{0}^{t} \hat{\mathcal{H}}(\tau)\,d\tau + \mathcal{O}(\Delta^2).
 \label{eq:C5}
\end{equation} Neglecting the $\mathcal{O}(\Delta^2)$-terms and replacing Eq.(\ref{eq:C5}) into Eq.(\ref{eq:C4}), the rate of transition can be expressed as
\begin{equation}
\begin{aligned}
 W &= \frac{d}{dt}  \left|\int_{0}^{t} \langle-| \hat{\mathcal{H}}(\tau) |+\rangle  d\tau   \right|^{2}  \equiv \lim_{t\rightarrow \infty} \frac{1 }{t}\left|\int_{0}^{t} \langle-| \hat{\mathcal{H}}(\tau) |+\rangle  d\tau   \right|^{2},
 \end{aligned}
 \label{eq:C6}
\end{equation}

Using Eq.(\ref{eq:C2})  and expanding the states $\{ |-\rangle,|+\rangle \}$ in terms of the manifold basis $\{ |g_{\text{a}}e_{\text{b}}\rangle, |e_{\text{a}}g_{\text{b}}\rangle \}$ as
\begin{equation}
\begin{aligned}
   |-\rangle &= \cos(\frac{\chi}{2})|g_{\textrm{a}} e_{\textrm{b}}\rangle+ \sin(\frac{\chi}{2})|e_{\textrm{a}} g_{\textrm{b}}\rangle,\\
   |+\rangle &= -\sin(\frac{\chi}{2})|g_{\textrm{a}} e_{\textrm{b}}\rangle+ \cos(\frac{\chi}{2})|e_{\textrm{a}} g_{\textrm{b}}\rangle,\\
 \end{aligned}
 \label{eq:C7}
\end{equation} with $\chi= \arctan(\frac{\Delta}{\varepsilon_{0} })$, the Eq.(\ref{eq:C6}) remains
\begin{equation}
\begin{aligned}
 W &= \lim_{t\rightarrow \infty}\frac{1 }{t}\left|\int_{0}^{t} \left(   \cos^{2}(\frac{\chi}{2})\Delta(\tau)  -  \sin^{2}(\frac{\chi}{2})\Delta(\tau)^{*} \right)  d\tau   \right|^{2}  ,\\
 &=\lim_{t\rightarrow \infty}\frac{1 }{t}  \left|  \cos^{2}(\frac{\chi}{2}) \int_{0}^{t}   \Delta(\tau)d\tau  -   \sin^{2}(\frac{\chi}{2})\int_{0}^{t}\Delta(\tau)^{*}  d\tau   \right|^{2}  .
 \end{aligned}
 \label{eq:C8}
\end{equation}

For simplicity, we are going to define the amount $\tilde{P}_{|a\rangle \rightarrow |b\rangle}(t) \in \mathbb{C}$ as
\begin{equation}
\begin{aligned}
 {P}_{|a\rangle \rightarrow |b\rangle}(t) &=  \left|\int_{0}^{t} \langle a | \hat{\mathcal{H}}(\tau) | b \rangle  d\tau   \right|^{2},\\
 &=  \left|   \tilde{P}_{|a\rangle \rightarrow |b\rangle}(t) \right|^{2},\\
  \tilde{P}_{|a\rangle \rightarrow |b\rangle}(t) &= \int_{0}^{t} \langle a | \hat{\mathcal{H}}(\tau) | b \rangle  d\tau.
 \end{aligned}
 \label{eq:C9}
\end{equation} Applying this definition to our case, we obtain
\begin{equation}
\begin{aligned}
 \tilde{P}_{|g_{\textrm{a}} e_{\textrm{b}}\rangle\rightarrow |e_{\textrm{a}} g_{\textrm{b}}\rangle}(t) &=\int_{0}^{t}   \Delta(\tau)d\tau,\\
 \tilde{P}_{|e_{\textrm{a}} g_{\textrm{b}}\rangle\rightarrow |g_{\textrm{a}} e_{\textrm{b}}\rangle}(t) &=\int_{0}^{t}   \Delta(\tau)^{*}d\tau.
 \end{aligned}
 \label{eq:C10}
\end{equation}

Replacing the Eq.(\ref{eq:C10}) into Eq.(\ref{eq:C8}), we obtain
\begin{equation}
\begin{aligned}
 W &=\lim_{t\rightarrow \infty}\frac{1 }{t}  \Big|  \cos^{2}(\frac{\chi}{2}) \tilde{P}_{|g_{\textrm{a}} e_{\textrm{b}}\rangle\rightarrow |e_{\textrm{a}} g_{\textrm{b}}\rangle}(t -   \sin^{2}(\frac{\chi}{2}) \tilde{P}_{|e_{\textrm{a}} g_{\textrm{b}}\rangle\rightarrow |g_{\textrm{a}} e_{\textrm{b}}\rangle}(t)   \Big|^{2}  .
  \end{aligned}
 \label{eq:C11}
\end{equation} Notice that the term $\left|\cos^{2}(\frac{\chi}{2}) \tilde{P}_{|g_{\textrm{a}} e_{\textrm{b}}\rangle\rightarrow |e_{\textrm{a}} g_{\textrm{b}}\rangle}   -   \sin^{2}(\frac{\chi}{2}) \tilde{P}_{|e_{\textrm{a}} g_{\textrm{b}}\rangle\rightarrow |g_{\textrm{a}} e_{\textrm{b}}\rangle}   \right|^{2}$ is similar to the calculation of the surviving probability when two different paths interfere that means $P_{s} = |A_{path, 1}-A_{path, 2}|^{2}$, with $A_{path, i} \in \mathbb{C}$ the quantum amplitude of each path. For our case, we can identify $A_{path, 1} \leftrightarrow \tilde{P}_{|g_{\textrm{a}} e_{\textrm{b}}\rangle\rightarrow |e_{\textrm{a}} g_{\textrm{b}}\rangle}(t)$ and $A_{path, 2} \leftrightarrow \tilde{P}_{|e_{\textrm{a}} g_{\textrm{b}}\rangle\rightarrow |g_{\textrm{a}} e_{\textrm{b}}\rangle}(t)$.

Now we proceed to expand the equation above, obtaining
\begin{equation}
\begin{aligned}
 W = \lim_{t\rightarrow \infty}\frac{1 }{t}  &\Big(  \cos^{4}(\frac{\chi}{2}) |\tilde{P}_{|g_{\textrm{a}} e_{\textrm{b}}\rangle\rightarrow (t)|e_{\textrm{a}} g_{\textrm{b}}\rangle}|^{2}   +  \sin^{4}(\frac{\chi}{2}) |\tilde{P}_{|e_{\textrm{a}} g_{\textrm{b}}\rangle\rightarrow |g_{\textrm{a}} e_{\textrm{b}}\rangle}(t)|^{2} \\
 & - 2 \cos^{2}(\frac{\chi}{2}) \sin^{2}(\frac{\chi}{2})Re [  \tilde{P}_{|g_{\textrm{a}} e_{\textrm{b}}\rangle\rightarrow |e_{\textrm{a}} g_{\textrm{b}}\rangle}(t)\tilde{P}^{*}_{|e_{\textrm{a}} g_{\textrm{b}}\rangle\rightarrow |g_{\textrm{a}} e_{\textrm{b}}\rangle} (t)  ] \Big)  .
  \end{aligned}
 \label{eq:C12}
\end{equation} Using the definitions presented in (\ref{eq:C9}) $W$ transforms to
\begin{equation}
\begin{aligned}
 W &= \lim_{t\rightarrow \infty}\frac{1 }{t}  \Big(  \cos^{4}(\frac{\chi}{2}) {P}_{|g_{\textrm{a}} e_{\textrm{b}}\rangle\rightarrow |e_{\textrm{a}} g_{\textrm{b}}\rangle}(t)   +  \sin^{4}(\frac{\chi}{2}) {P}_{|e_{\textrm{a}} g_{\textrm{b}}\rangle\rightarrow |g_{\textrm{a}} e_{\textrm{b}}\rangle}(t)  \\
 &\hspace {1.5cm}- 2 \cos^{2}(\frac{\chi}{2}) \sin^{2}(\frac{\chi}{2})Re [  \tilde{P}_{|g_{\textrm{a}} e_{\textrm{b}}\rangle\rightarrow |e_{\textrm{a}} g_{\textrm{b}}\rangle}(t)\tilde{P}^{*}_{|e_{\textrm{a}} g_{\textrm{b}}\rangle\rightarrow |g_{\textrm{a}} e_{\textrm{b}}\rangle} (t)  ] \Big)  ,\\
 &= \cos^{4}(\frac{\chi}{2}) \lim_{t\rightarrow \infty}\frac{ {P}_{|g_{\textrm{a}} e_{\textrm{b}}\rangle\rightarrow |e_{\textrm{a}} g_{\textrm{b}}\rangle}(t) }{t}   +  \sin^{4}(\frac{\chi}{2}) \lim_{t\rightarrow \infty}\frac{ {P}_{|e_{\textrm{a}} g_{\textrm{b}}\rangle\rightarrow |g_{\textrm{a}} e_{\textrm{b}}\rangle} (t)}{t}  \\
 & - 2 \cos^{2}(\frac{\chi}{2}) \sin^{2}(\frac{\chi}{2})\lim_{t\rightarrow \infty}\frac{ Re [  \tilde{P}_{|g_{\textrm{a}} e_{\textrm{b}}\rangle\rightarrow |e_{\textrm{a}} g_{\textrm{b}}\rangle} (t)\tilde{P}^{*}_{|e_{\textrm{a}} g_{\textrm{b}}\rangle\rightarrow |g_{\textrm{a}} e_{\textrm{b}}\rangle}(t)}{t}   ]  ,\\
  &= \cos^{4}(\frac{\chi}{2}) W_{|g_{\textrm{a}} e_{\textrm{b}}\rangle\rightarrow |e_{\textrm{a}} g_{\textrm{b}}\rangle}  +  \sin^{4}(\frac{\chi}{2}){W}_{|e_{\textrm{a}} g_{\textrm{b}}\rangle\rightarrow |g_{\textrm{a}} e_{\textrm{b}}\rangle}  \\
  & - 2 \cos^{2}(\frac{\chi}{2}) \sin^{2}(\frac{\chi}{2})\lim_{t\rightarrow \infty}\frac{1 }{t}Re [  \tilde{P}_{|g_{\textrm{a}} e_{\textrm{b}}\rangle\rightarrow |e_{\textrm{a}} g_{\textrm{b}}\rangle}\tilde{P}^{*}_{|e_{\textrm{a}} g_{\textrm{b}}\rangle\rightarrow |g_{\textrm{a}} e_{\textrm{b}}\rangle}   ].\\
  \end{aligned}
 \label{eq:C13}
\end{equation} This last result shows how the transition rate depends on the individual transition rates  $W_{|a\rangle \rightarrow |b\rangle}$  plus a correction given by the quantum interference between the states. Moreover, each term is normalized by a factor that depends on how the system was initially prepared. As it was mentioned before, this result is similar to the expanded expression $ P_{s}=  |A_{path,1}|^2 + |A_{path,2}|^2 - Re \left[A^{*}_{path,1}A_{path,2}\right] $. Since $P_{s}$ is calculated in terms of  quantum amplitudes, the  final $P_{s}$ expression presents a \textit{classical} counterpart, linked with the qubit path probabilities $P_{s}=|A_{path,i}|^2 $, along with a \textit{quantum} counterpart, linked with the interference term $Re \left[A^{*}_{path,1}A_{path,2}\right]$. Analogously with the disorder systems, the interference term survives disorder averaging when the system presents time-reversal symmetry. In this way,  the term $\langle Re \left[A^{*}_{path,1}A_{path,2}\right] \rangle$ depicts a negative correction to the surviving probability $\langle P_{s} \rangle$, i.e. the transition probability $\langle W \rangle$, with $\langle ... \rangle$ representing the disorder averaging.

We must still calculate the unknown quantities $ W_{|g_{\textrm{a}} e_{\textrm{b}}\rangle\rightarrow |e_{\textrm{a}} g_{\textrm{b}}\rangle}$, $ W_{|e_{\textrm{a}} g_{\textrm{b}}\rangle\rightarrow |g_{\textrm{a}} e_{\textrm{b}}\rangle}$ and $Re [  \tilde{P}_{|g_{\textrm{a}} e_{\textrm{b}}\rangle\rightarrow |e_{\textrm{a}} g_{\textrm{b}}\rangle}(t) \\ \tilde{P}^{*}_{|e_{\textrm{a}} g_{\textrm{b}}\rangle\rightarrow |g_{\textrm{a}} e_{\textrm{b}}\rangle}(t)   ]  $ in terms of the driving parameters.

The first step is to calculate $\Delta(t)=\Delta e^{-i\phi(t)}$, to this end, we can use the Bessel function  property $e^{i x\sin(\theta)} = \sum_{n} J_{n}(x)e^{in\theta}$, $n\in \mathbb{Z}$, obtaining:
\begin{equation}
\begin{aligned}
	\Delta(t) = \Delta \sum_{n m } J_{n}\left(\frac{A_{1}}{\omega_{1}}\right)J_{m}\left(\frac{A_{2}}{\omega_{2}}\right) e^{i(\varepsilon_{0}+n\omega_{1} + m \omega_{2})t}e^{im\alpha}.
  \end{aligned}
 \label{eq:C14}
\end{equation}

Replacing the Eq.(\ref{eq:C14}) into  Eqs.(\ref{eq:C9}) and  (\ref{eq:C10}), it follows
\begin{equation}
\begin{aligned}
 {P}_{|g_{\textrm{a}} e_{\textrm{b}}\rangle\rightarrow |e_{\textrm{a}} g_{\textrm{b}}\rangle}(t) &=\frac{\Delta^{2}}{4} \sum_{nmn'm'} J_{n}\left(\frac{A_{1}}{\omega_{1}}\right)J_{m}\left(\frac{A_{2}}{\omega_{2}}\right)J_{n'}\left(\frac{A_{1}}{\omega_{1}}\right)J_{m'}\left(\frac{A_{2}}{\omega_{2}}\right) e^{im\alpha}e^{-im'\alpha}\\
&e^{i((n-n')\omega_{1} + (m-m') \omega_{2})t/2}\frac{\sin(\varepsilon_{0}+ n\omega_{1} + m \omega_{2})t/2}{(\varepsilon_{0}+ n\omega_{1} + m \omega_{2})/2}\frac{\sin(\varepsilon_{0}+ n'\omega_{1} + m' \omega_{2})t/2}{(\varepsilon_{0}+ n'\omega_{1} + m' \omega_{2})/2}.
 \end{aligned}
 \label{eq:C15}
\end{equation}

For another part, the interference term can be written as
\begin{equation}
\begin{aligned}
&Re\Big[ \tilde{P}_{|g_{\textrm{a}} e_{\textrm{b}}\rangle\rightarrow |e_{\textrm{a}} g_{\textrm{b}}\rangle}(t) \tilde{P}^{*}_{|e_{\textrm{a}} g_{\textrm{b}}\rangle\rightarrow |g_{\textrm{a}} e_{\textrm{b}}\rangle}(t) \Big]=\\
& \frac{\Delta^{2}}{4}  Re\Big[\sum_{nmn'm'} J_{n}\left(\frac{A_{1}}{\omega_{1}}\right)J_{m}\left(\frac{A_{2}}{\omega_{2}}\right)J_{n'}\left(\frac{A_{1}}{\omega_{1}}\right)J_{m'}\left(\frac{A_{2}}{\omega_{2}}\right) e^{im\alpha}e^{im'\alpha} e^{i((n+n')\omega_{1} + (m+m') \omega_{2})t/2}\\
& \hspace{1.5cm}\frac{\sin(\varepsilon_{0}+ n\omega_{1} + m \omega_{2})t/2}{(\varepsilon_{0}+ n\omega_{1} + m \omega_{2})/2}\frac{\sin(\varepsilon_{0}+ n'\omega_{1} + m' \omega_{2})t/2}{(\varepsilon_{0}+ n'\omega_{1} + m' \omega_{2})/2} \Big].
 \end{aligned}
 \label{eq:C16}
\end{equation}

At this point, it should be noted that the calculations above are different from \cite{3,4}, since in our case the frequency  $\omega$ of driving is small compared to the energy gap $\Delta$, thus  it is not possible to neglect the fast oscillating terms. In this way, the final calculations are slightly different.

For simplicity, we are going to considering the limit when $\alpha \rightarrow 0$, thus $e^{i m\alpha} \approx 1 + im\alpha$. From Eq.(\ref{eq:C15}) we obtain
\begin{equation}
\begin{aligned}
 {P}_{|g_{\textrm{a}} e_{\textrm{b}}\rangle\rightarrow |e_{\textrm{a}} g_{\textrm{b}}\rangle}(t) &\approx \frac{\Delta^{2}}{4} \Big|\sum_{nm} J_{n}\left(\frac{A_{1}}{\omega_{1}}\right)J_{m}\left(\frac{A_{2}}{\omega_{2}}\right) e^{i((n\omega_{1} + m \omega_{2})t/2}\frac{\sin(\varepsilon_{0}+ n\omega_{1} + m \omega_{2})t/2}{(\varepsilon_{0}+ n\omega_{1} + m \omega_{2})/2} \Big|^{2}\\
&+ \frac{\Delta^{2}}{4} \alpha^{2}\Big|\sum_{nm} m J_{n}\left(\frac{A_{1}}{\omega_{1}}\right)J_{m}\left(\frac{A_{2}}{\omega_{2}}\right)e^{i((n\omega_{1} + m \omega_{2})t/2}\frac{\sin(\varepsilon_{0}+ n\omega_{1} + m \omega_{2})t/2}{(\varepsilon_{0}+ n\omega_{1} + m \omega_{2})/2} \Big|^{2}\\
&+ \frac{\Delta^{2}}{4} i\alpha  \sum_{nmn'm'} (m-m')J_{n}\left(\frac{A_{1}}{\omega_{1}}\right)J_{m}\left(\frac{A_{2}}{\omega_{2}}\right)J_{n'}\left(\frac{A_{1}}{\omega_{1}}\right)J_{m'}\left(\frac{A_{2}}{\omega_{2}}\right)e^{i((n-n')\omega_{1} + (m-m') \omega_{2})t/2}\\
&\frac{\sin(\varepsilon_{0}+ n\omega_{1} + m \omega_{2})t/2}{(\varepsilon_{0}+ n\omega_{1} + m \omega_{2})/2}\frac{\sin(\varepsilon_{0}+ n'\omega_{1} + m' \omega_{2})t/2}{(\varepsilon_{0}+ n'\omega_{1} + m' \omega_{2})/2}.
 \end{aligned}
 \label{eq:C17}
\end{equation}  Taking the limit  $\lim_{t\rightarrow \infty}\frac{{P}_{|g_{\textrm{a}} e_{\textrm{b}}\rangle\rightarrow |e_{\textrm{a}} g_{\textrm{b}}\rangle}(t)  }{t}$ and using the well known result $\lim_{t\rightarrow \infty}\frac{1}{t} \frac{\sin^{2} (\beta t )}{\beta^2} \rightarrow 2\pi \delta(\beta)$, the corresponding transition rate can be approximately calculated as
\begin{equation}
\begin{aligned}
 {W}_{|g_{\textrm{a}} e_{\textrm{b}}\rangle\rightarrow |e_{\textrm{a}} g_{\textrm{b}}\rangle} &= \lim_{t\rightarrow \infty}\frac{{P}_{|g_{\textrm{a}} e_{\textrm{b}}\rangle\rightarrow |e_{\textrm{a}} g_{\textrm{b}}\rangle}(t)  }{t}\\
&\approx \frac{\Delta^{2}}{4} \sum_{nm} J_{n}\left(\frac{A_{1}}{\omega_{1}}\right)^{2}J_{m}\left(\frac{A_{2}}{\omega_{2}}\right)^{2} \delta(\varepsilon_{0} + n\omega_{1} + m\omega_{2})\\
&+ \frac{\Delta^{2}}{4} \alpha^{2} \sum_{nm} m^{2}J_{n}\left(\frac{A_{1}}{\omega_{1}}\right)^{2}J_{m}\left(\frac{A_{2}}{\omega_{2}}\right)^{2} \delta(\varepsilon_{0} + n\omega_{1} + m\omega_{2}) \\
&\approx {W}_{|g_{\textrm{a}} e_{\textrm{b}}\rangle\rightarrow |e_{\textrm{a}} g_{\textrm{b}}\rangle}^{\alpha = 0} + \alpha^{2} {\xi}_{|g_{\textrm{a}} e_{\textrm{b}}\rangle\rightarrow |e_{\textrm{a}} g_{\textrm{b}}\rangle}^{\alpha \neq 0}
 \end{aligned}
 \label{eq:C18}
\end{equation}

Using the same procedure presented above to calculate the interference term, we obtain
\begin{equation}
    \begin{aligned}
    & \lim_{t\rightarrow \infty}\frac{1 }{t}Re\Big[ \tilde{P}_{|g_{\textrm{a}} e_{\textrm{b}}\rangle\rightarrow |e_{\textrm{a}} g_{\textrm{b}}\rangle}(t) \tilde{P}^{*}_{|e_{\textrm{a}} g_{\textrm{b}}\rangle\rightarrow |g_{\textrm{a}} e_{\textrm{b}}\rangle}(t) \Big] \approx\\
    &\approx \frac{\Delta^{2}}{4} \lim_{t\rightarrow \infty}\frac{1 }{t}Re\Big[ \left( \sum_{nm} J_{n}\left(\frac{A_{1}}{\omega_{1}}\right)J_{m}\left(\frac{A_{2}}{\omega_{2}}\right) e^{i((n\omega_{1} + m \omega_{2})t/2}\frac{\sin(\varepsilon_{0}+ n\omega_{1} + m \omega_{2})t/2}{(\varepsilon_{0}+ n\omega_{1} + m \omega_{2})/2} \right)^{2} \Big]\\
    &- \frac{\Delta^{2}}{4} \alpha^{2}\lim_{t\rightarrow \infty}\frac{1 }{t}Re \Big[ \left( \sum_{nm} mJ_{n}\left(\frac{A_{1}}{\omega_1}\right)
    J_{m} \left( \frac{A_{2}}{\omega_{2}} \right) e^{i(n\omega_1 + m \omega_2)t/2} \frac{\sin(\varepsilon_{0}+ n\omega_{1} + m \omega_{2})t/2}{(\varepsilon_{0}+ n\omega_{1} + m \omega_{2})/2} \right)^{2} \Big] \\
    &\approx \frac{\Delta^{2}}{4} \sum_{nm} J_{n}\left(\frac{A_{1}}{\omega_{1}}\right)^{2}J_{m}\left(\frac{A_{2}}{\omega_{2}}\right)^{2} \delta(\varepsilon_{0} + n\omega_{1} + m\omega_{2}) \\
    &- \frac{\Delta^{2}}{4} \alpha^{2}\sum_{nm} m^{2}J_{n}\left(\frac{A_{1}}{\omega_{1}}\right)^{2}J_{m}\left(\frac{A_{2}}{\omega_{2}}\right)^{2} \delta(\varepsilon_{0} + n\omega_{1} + m\omega_{2}) \\
    &\approx \lim_{t\rightarrow \infty}\frac{1 }{t}Re\Big[ \tilde{P}_{|g_{\textrm{a}} e_{\textrm{b}}\rangle\rightarrow |e_{\textrm{a}} g_{\textrm{b}}\rangle}(t) \tilde{P}^{*}_{|e_{\textrm{a}} g_{\textrm{b}}\rangle\rightarrow |g_{\textrm{a}} e_{\textrm{b}}\rangle}(t) \Big]^{\alpha = 0} - \alpha^{2} \eta^{\alpha \neq 0}.
    \end{aligned}
 \label{eq:C19}
\end{equation}

Finally, from Eqs.(\ref{eq:C18})  and (\ref{eq:C19}) we can obtain an approximated equation for the total transition rate around the point $\alpha \approx 0$, which reads
\begin{equation}
\begin{aligned}
 W&\approx \cos^{4}(\frac{\chi}{2}) W_{|g_{\textrm{a}} e_{\textrm{b}}\rangle\rightarrow |e_{\textrm{a}} g_{\textrm{b}}\rangle}^{\alpha=0}  +  \sin^{4}(\frac{\chi}{2}){W}_{|e_{\textrm{a}} g_{\textrm{b}}\rangle\rightarrow |g_{\textrm{a}} e_{\textrm{b}}\rangle}^{\alpha=0}  \\
 &- 2 \cos^{2}(\frac{\chi}{2}) \sin^{2}(\frac{\chi}{2})\lim_{t\rightarrow \infty}\frac{1 }{t}Re [  \tilde{P}_{|g_{\textrm{a}} e_{\textrm{b}}\rangle\rightarrow |e_{\textrm{a}} g_{\textrm{b}}\rangle}\tilde{P}^{*}_{|e_{\textrm{a}} g_{\textrm{b}}\rangle\rightarrow |g_{\textrm{a}} e_{\textrm{b}}\rangle}   ]^{\alpha=0} \\
 &+ \alpha^{2} \left( \cos^{4}(\frac{\chi}{2}) \xi_{|g_{\textrm{a}} e_{\textrm{b}}\rangle\rightarrow |e_{\textrm{a}} g_{\textrm{b}}\rangle}^{\alpha\neq0}  +  \sin^{4}(\frac{\chi}{2})\xi_{|e_{\textrm{a}} g_{\textrm{b}}\rangle\rightarrow |g_{\textrm{a}} e_{\textrm{b}}\rangle}^{\alpha\neq0}  + 2 \cos^{2}(\frac{\chi}{2}) \sin^{2}(\frac{\chi}{2})  \eta^{\alpha \neq 0} \right),\\
&\approx  W_{\alpha=0} + \alpha^{2} \zeta_{\alpha\neq0}
  \end{aligned}
 \label{eq:C20}
 \end{equation}

 Applying the average over initial conditions to the Eq.(\ref{eq:C20}), we obtain the general equation
 \begin{equation}
\begin{aligned}
 \langle W \rangle &\approx   \langle W \rangle_{\alpha=0}  +  \alpha^{2}\langle \zeta \rangle_{\alpha\neq0}\\
 \langle W \rangle &-  \langle W \rangle_{\alpha=0} \approx \alpha^{2} \langle \zeta \rangle_{\alpha\neq0},
  \end{aligned}
 \label{eq:C21}
 \end{equation} defining the averaged transition rate as $\langle W \rangle = \frac{1}{|\varepsilon_{0,max} - \varepsilon_{0,min}|} \int_{\varepsilon_{0,min}}^{\varepsilon_{0,max}} W \, d\varepsilon_{0}$ and wherewith $\langle W \rangle_{\alpha=0}$ and $\langle \zeta \rangle_{\alpha \neq 0}$ are positive quantitites. From this last result, we can conclude that the observation of weak localization in this system is possible.

\section{\label{sec:D} Quantum simulator and  Random Matrix theory predictions for  disordered mesoscopic systems }

In this section we review  the WL and UCF  predicted by
Radom Matrix Theory \cite{9} for disordered  mesoscopic systems,  with the aim to compare with our results obtained for the quantum simulator.
\begin{figure*}[ht!]
\centering
	\includegraphics[width = 6.5 cm]{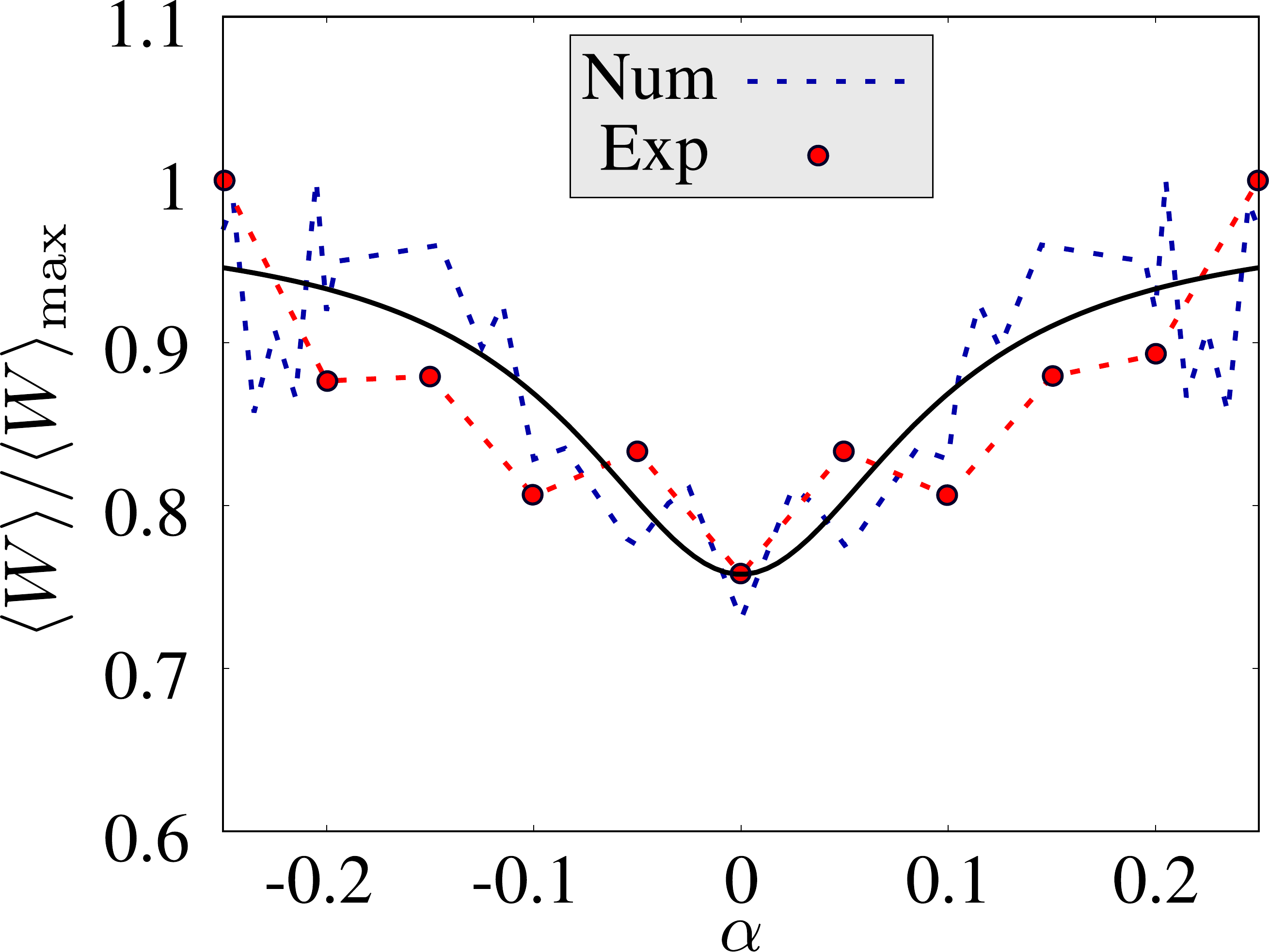}
	\caption{\textbf{First-order statistic the transition rate: \text{WL}}  Experimental and numerical results of the normalized transition rate $\langle W \rangle/\langle W \rangle_{\textrm{max}}$ as a function of the asymmetry parameter $\alpha$.  $\langle W \rangle$ is the transition rate ensemble-averaged over all $f_{dc}$ values for fixed $\alpha$, and $\langle W \rangle_{\textrm{max}}$ the corresponding maximum value for each case. The fitting curve using Eq.\eqref{eq:W_alpha} in plotted in bold line.}
\label{fig:2-suppl}
\end{figure*}

By assuming a diffusive (or fully chaotic billiard) transport regime \cite{5}, the suppression of the WL correction in the average conductance $\langle G \rangle $ follows a Lorentzian shape :
\begin{equation}
    \begin{aligned}
    	 \langle G \rangle(B)&= a - \frac{b}{1 + \left(\frac{B}{B_{c}}\right)^2},
	  \label{eq:W_B}
    \end{aligned}
\end{equation} with $B$ the external magnetic field, $B_{c}$ the critical magnetic field, and  $a,b\in\mathbb{R}$ parameters depending on the system characteristics.
In our case, we have obtained that  the average transition rate  $\langle W \rangle$ satisfies
\begin{equation}
    \begin{aligned}
    	 \langle  W \rangle_{\alpha} &= \tilde{a} - \frac{\tilde{b}}{1 + \left(\frac{\alpha}{\alpha_{c}}\right)^2},
	  \label{eq:W_alpha}
    \end{aligned}
\end{equation} with $\alpha$  the time reversal breaking control parameter and  $\alpha_{c}$ the critical parameter playing the role of the  critical magnetic field in Eq.\ref{eq:W_B}. In this last case $\tilde{a}$ and $\tilde{b}\in\mathbb{R}$  are constants depending on the properties of the quantum simulator. Fitting Eq.\eqref{eq:W_alpha} with the numerical results, we have  obtained   $\alpha_c = (0.09 \pm 0.03)$. Fig.\ref{fig:2-suppl} displays  the fitting curve along with the numerical and experimental results, showing a good agreement.

In particular, performing a Taylor expansion of Eq.\eqref{eq:W_alpha} around $\alpha \sim 0$, we get
\begin{equation}
    \begin{aligned}
    	 \langle W \rangle_{\alpha} & \sim  \tilde{c} + \frac{b}{\alpha_{c}^2}\alpha^2,
	  \label{eq:W_alpha_Taylor}
    \end{aligned}
\end{equation} with $\tilde{c}= \tilde{a} - \tilde{b}$. Notice that the Eq.\eqref{eq:W_alpha_Taylor} is similar to the analytical expression obtained in Eq.\eqref{eq:C21}.

It should be stressed that  as our quantum simulator operates in the limit of  few scattering centers,  the RMT predictions valid for highly  disordered transport  regime are not necessarily  fulfilled.  In particular in  Ref.\onlinecite{6} it has been shown  that the WL correction can follow a dependence of the form $|B|$ for a  non- fully chaotic regime.

\begin{figure*}[ht!]
\centering
	\hspace{3.5cm}\includegraphics[width = 10.2 cm]{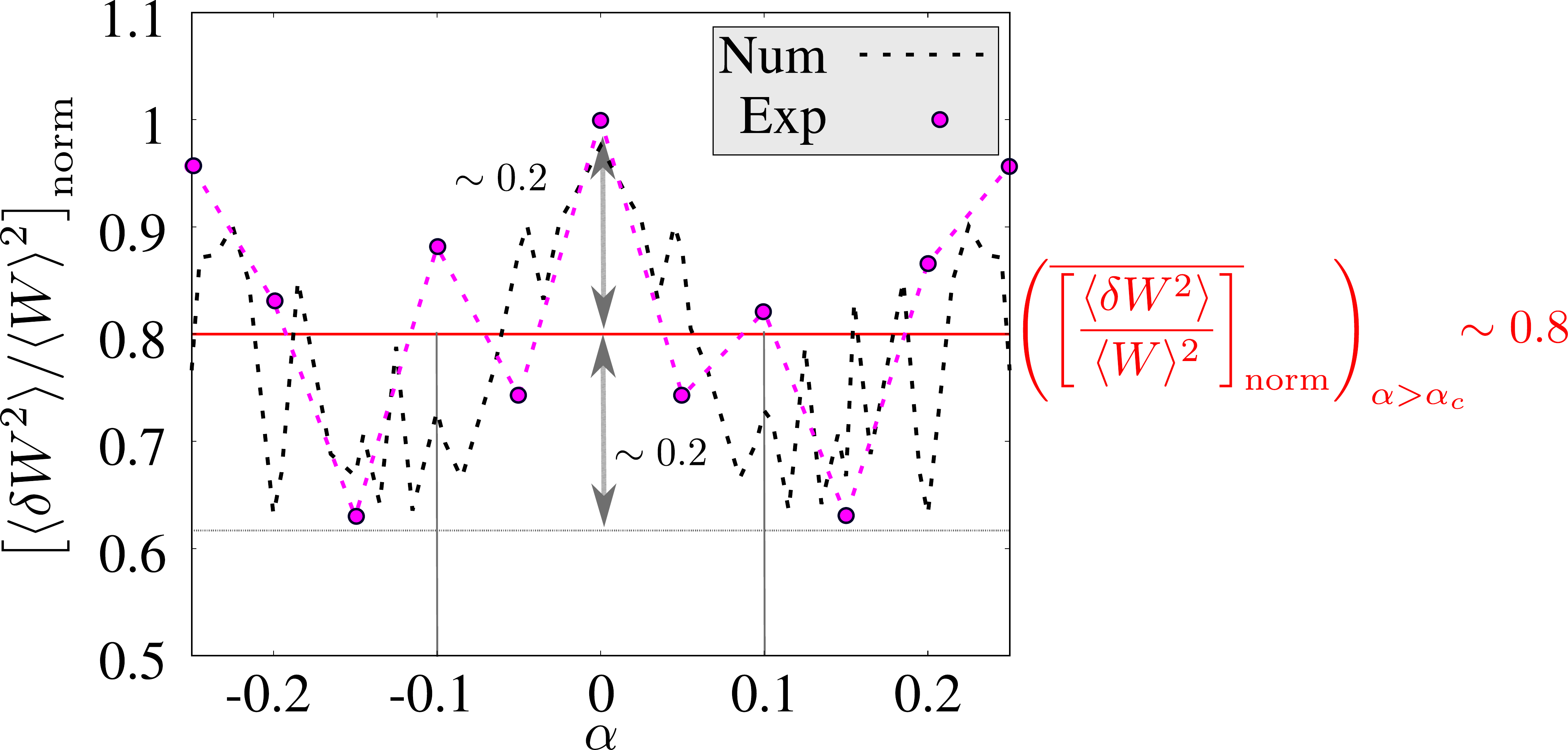}
	\caption{\textbf{Second-order statistic of the transition rate: UCF.} Experimental and numerical results for the normalized variance $[\langle \delta W^2 \rangle / \langle W \rangle^2 ]_\text{norm}$. The average of the fluctuations when $\alpha \neq 0 $ is plotted in bold red line, corresponding to the amount $\left( \overline{[\langle \delta W^{2}\rangle/\langle W\rangle^{2}]}_{\text{norm}} \right)_{\alpha>\alpha_{c}}\sim 0.8$ for both numerical and experimental results, see Eq.\eqref{eq:UCF_average}.}
\label{fig:3-suppl}
\end{figure*}

The RMT  predictions for the COE (time reversal symmetry (TRS)) and CUE (non-TRS)  ensembles \cite{7,8} satisfy $ \langle \delta G^{2} \rangle_{\text{TRS}} \sim 2 \langle \delta G^{2} \rangle_{\text{non-TRS}}$, with $\text{var}(G)=\langle \delta G^{2} \rangle$.
In our case, in order to perform a similar comparison, we have defined the average fluctuation $\overline{[\langle \delta W^{2}\rangle/\langle W \rangle^{2}]}_{\text{norm}}$ for different  $\alpha$ values satisfying   $\alpha \neq > \alpha_{c}$ in order to consider the non-TRS case, and after  fitting with Eq.\eqref{eq:W_alpha} we obtained :
\begin{equation}
    \begin{aligned}
    	 \left( \overline{\left[\frac{\langle \delta W^{2}\rangle}{\langle W\rangle^{2}}\right] }_{\text{norm}}  \right)_{\alpha>\alpha_{c}}  &\sim 0.8139 \text{ , experimental results}\\
	     \left( \overline{\left[\frac{\langle \delta W^{2}\rangle}{\langle W\rangle^{2}}\right] }_{\text{norm}}  \right)_{\alpha>\alpha_{c}} &\sim 0.8136 \text{ , numerical results}.
	  \label{eq:UCF_average}
    \end{aligned}
\end{equation}

In Fig.\ref{fig:3-suppl} we display the experimental and numerical results of the UCF as a function of $\alpha$. The  value  $\left( \overline{[\langle \delta W^{2}\rangle/\langle W\rangle^{2}]}_{\text{norm}} \right)_{\alpha>\alpha_{c}}\sim 0.8$ is plotted in red bold line. Notice that  the ratio $\langle \delta W ^2 \rangle_{\alpha=0}/\langle \delta W ^2 \rangle_{\alpha>\alpha_{c}} \sim 2$ is satisfied for both numerical and experimental results. The arrows schematically plotted in Fig.\ref{fig:3-suppl} show that the averaged fluctuations and the UCF peak are measured from the minimum value of $[\langle \delta W^2 \rangle / \langle W \rangle^2 ]_\text{norm}$ instead of from  zero.

In the case of COE ensemble \cite{9}, the WL and UCF satify respectively:
\begin{equation}
    \begin{aligned}
    	  \langle G \rangle_{\text{TRS}} &= -\frac{2}{3} \left( \frac{e^2}{ h} \right) ,\\
	      \langle \delta G^{2} \rangle_{\text{TRS}}&= \frac{2}{15} \left( \frac{e^2}{ h} \right)^2.
	  \label{eq:diffusive_wire}
    \end{aligned}
\end{equation} In order to compare these typical values with our results we define the dimensionless ratio:
\begin{equation}
    \begin{aligned}
    	 r_{\text{RMT}}=\left(\frac{ \langle \delta G^{2} \rangle}{\langle G \rangle^2}\right)_{\text{TRS}} &= \frac{3}{10} \sim 0.3.
	  \label{eq:ratio_diff_wires}
    \end{aligned}
\end{equation} In our case, taking into account the previous results, the ratio gives
\begin{equation}
    \begin{aligned}
    	    r=\left(\frac{ \langle \delta W^{2} \rangle}{\langle W \rangle^2}\right)_{\alpha=0} &\sim 0.5585 \text{ , experimental results} \\
	        r=\left(\frac{ \langle \delta W^{2} \rangle}{\langle W \rangle^2}\right)_{\alpha=0} &\sim 0.6958 \text{ , numerical results},
	  \label{eq:ratio_our_case}
    \end{aligned}
\end{equation}  being $r$ of the order of the $r_{\text{RMT}}$.

\def\section#1#2{\oldsection#1#2}

\end{document}